\newcommand{\CLASSINPUTbottomtextmargin}{0.7in}
\documentclass[journal]{IEEEtran}

\usepackage{cite}
\usepackage[utf8]{inputenc}
\usepackage{amsmath,amssymb,amsfonts}
\usepackage{algorithmic}
\usepackage{graphicx}
\usepackage{textcomp}
\usepackage{url}
\usepackage[caption=false]{subfig}
\usepackage{tabu}
\usepackage{booktabs}
\usepackage{siunitx}
\usepackage{xcolor}
\usepackage{soulutf8}
\usepackage{paralist}

\usepackage[switch]{lineno}
\linenumbers

\renewcommand\hl[1]{#1}

\usepackage{tikz}

\newcommand\copyrighttext{%
  \footnotesize \textcopyright~2020 IEEE. Personal use of this material is permitted. Permission from IEEE must be obtained for all other uses, in any current or future media, including reprinting/republishing this material for advertising or promotional purposes, creating new collective works, for resale or redistribution to servers or lists, or reuse of any copyrighted component of this work in other works. DOI: \texttt{10.1109/TNSM.2020.3033071}, IEEE}
\newcommand\copyrightnotice{%
\begin{tikzpicture}[remember picture,overlay]
\node[anchor=south,yshift=10pt] at (current page.south) {\fbox{\parbox{\dimexpr\textwidth-\fboxsep-\fboxrule\relax}{\copyrighttext}}};
\end{tikzpicture}%
}

 \DeclareSIUnit{\bit}{b}
 \DeclareSIUnit{\byte}{B}
 
\usepackage[final]{changes}
\definechangesauthor[name={Felipe}, color=cyan]{FG}
\definechangesauthor[name={Javier}, color=purple]{FJ}

\setdeletedmarkup{{\color{red} \st{#1}}}
\setauthormarkup{}

\usepackage{mdframed}
\definecolor{light-gray}{gray}{0.95} 
\global\mdfdefinestyle{exampledefault}{%
backgroundcolor=light-gray, roundcorner=10pt,leftmargin=1, rightmargin=1, innerleftmargin=15, innertopmargin=15,innerbottommargin=15, outerlinewidth=1, linecolor=light-gray}

\begin{document}

\title{A Software-Defined Networking Solution for Transparent Session and Service Continuity in Dynamic Multi-Access Edge Computing}
\author{%
  \IEEEauthorblockN{Pablo Fondo-Ferreiro, Felipe Gil-Castiñeira, Francisco Javier González-Castaño, David Candal-Ventureira}
  
  \IEEEauthorblockA{atlanTTic Research Center, University of Vigo\\
    E.E. Telecomunicación, Rúa Maxwell s/n, 36310 Vigo, Spain.\\
    Tel.:+34~986~818684; email:~\url{{pfondo, xil, javier, dcandal}@gti.uvigo.es}
}%
}

\markboth
{DRAFT}{Fondo-Ferreiro \headeretal: An SDN Solution for Transparent Session and Service Continuity in Dynamic MEC}

\maketitle
\copyrightnotice

\begin{abstract}
Multi-Access Edge Computing (MEC) 
will allow implementing low-latency services that have been unfeasible so far. 
The European Telecommunications Standards Institute (ETSI) and the 3rd Generation Partnership Project (3GPP) are working towards the standardization of MEC in 5G networks and the corresponding solutions for routing user traffic to  applications in local area networks.
Nevertheless, there are neither practical implementations
for dynamically relocating applications from the core to a MEC host nor from one MEC host to another ensuring service continuity.
In this paper we propose a solution based on  Software-Defined Networking (SDN) to create a new instance of the IP anchor point to dynamically redirect User Equipment (UE) traffic to a new physical location (e.g. an edge infrastructure).
We also present a novel approach that leverages  SDN 
to replicate the previous context of the connection in the new instance of the IP anchor point, thus guaranteeing Session and Service Continuity (SSC), and compare 
it with alternative state replication strategies. This approach can be used to implement edge services in 4G or 5G networks. 
\end{abstract}


\IEEEpeerreviewmaketitle


\section{Introduction}
\label{sec:introduction}

\IEEEPARstart{5}{G} flexibility, agility, and network and context awareness will enhance end-user Quality of Experience (QoE) \cite{hu2015mobile}. Virtualization and cloud computing allow building flexible and efficient architectures. However, traditional application deployment in distant datacenters is limiting. The cloud was not  designed for stringent latency constraints, and therefore it 
cannot support 5G network-side applications such as high-quality Augmented Reality (AR) image generation  and low-latency remote gaming. Hence the paradigms to take them to the network edge such as Fog Computing, cloudlets, and MEC. 

MEC is specially relevant in 5G scenarios.  In 2014, the ETSI launched the Multi-Access Edge Computing Industry Specification Group (MEC ISG) to standardize the integration of third party applications into multi-vendor edge platforms \cite{giust2017multi}. This group considered that MEC components and Mobile Edge (ME) applications will be implemented on top of the  Network Functions Virtualization (NFV) architecture so they will share the same infrastructure and management functions. This will allow operators to maximize return from virtualization investments  \cite{etsi017} and provide new edge services.

In \cite{giust2018mec} the ETSI studied  MEC systems for 4G architectures, and identified the challenge  to route  IP packets to MEC applications in a distributed architecture.
The components involved must be MEC-aware  to manage the session and correctly route the traffic for each UE, and the MEC needs to be aware of UE mobility to maintain service continuity even 
during MEC handovers.  ETSI also points out the open issue of relocating a mobile edge application between a cloud environment and a MEC host.  In this paper we propose a solution to  this problem that does not require the MEC components to be aware of UE mobility. Applications may be dynamically relocated between the cloud and the edge, for instance to take them physically closer to the user.  

Consider for example a cloud gaming case in which a low end device does not have enough graphic computing power. A cloud-based game rendering service might be relocated to the edge to follow the user, thus minimizing the latency between user interactions and  presentation of the rendered image. Or consider the case of user application offloading with varying performance requirements. 
However, edge resources are limited~\cite{tran2017collaborative} and thus placing applications permanently there is unfeasible. Alternatively, applications can be moved between the edge and the cloud depending on trade-offs between computing costs and latency requirements.
Ideally,  application relocation should not interrupt ongoing user sessions. Our solution achieves this goal.

\begin{figure}[ht!]
\centering
\includegraphics[width=0.99\columnwidth]{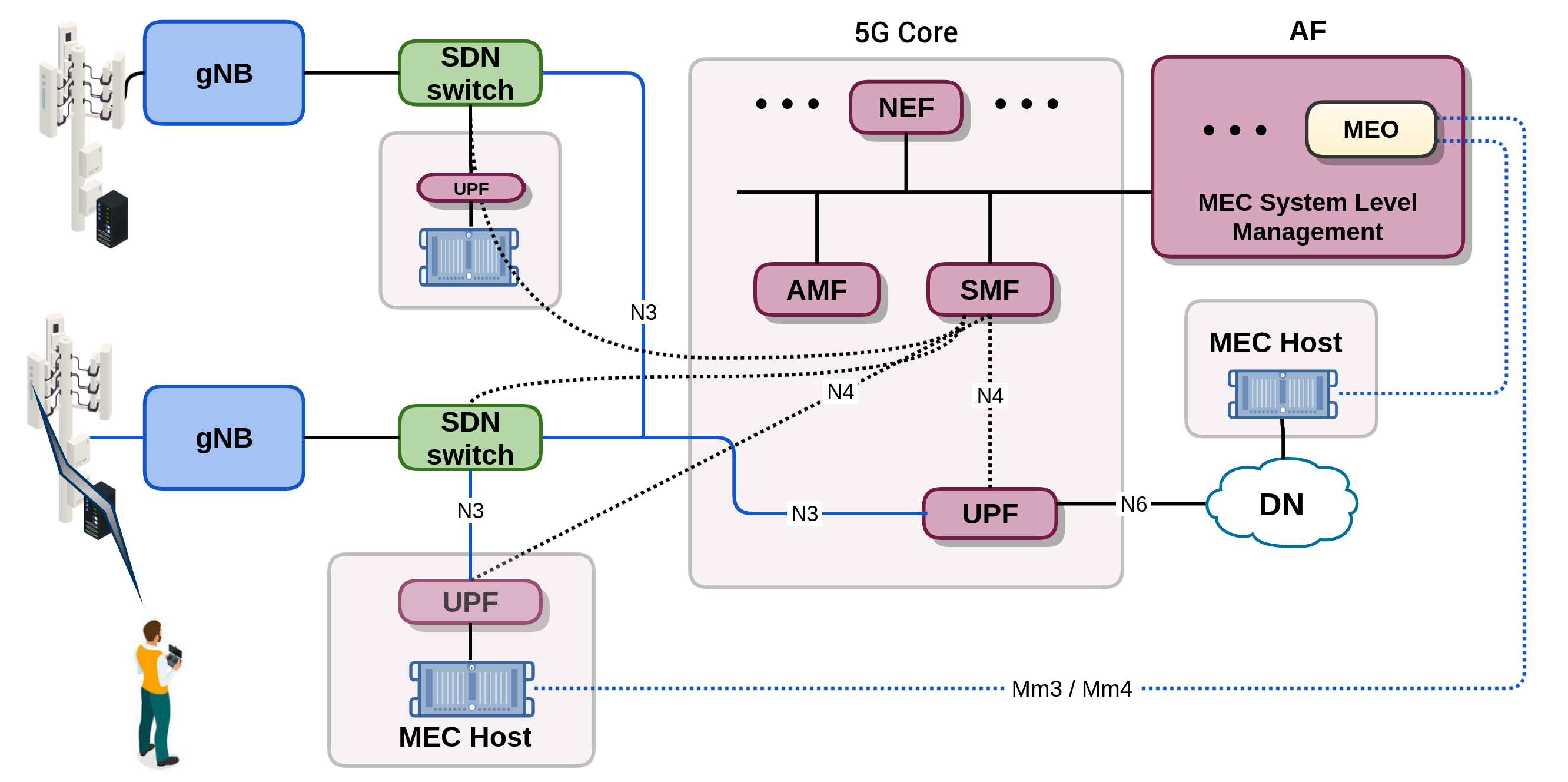}
\caption{Simplified view of the integration of our proposal in 3GPP 5G and ETSI MEC architectures}
\label{fig:5gintegration}
\end{figure}

Our proposal is also relevant for 5G networks, where MEC has been identified as one of the key technologies for supporting low latency. Even though the Service Based Architecture (SBA) facilitates the deployment and integration of MEC, it is still necessary to propose solutions to direct the traffic from a particular moving UE to the serving application and to take network-side applications and their traffic to the nearest MEC node \cite{Kekki2018mec}. In  5G  networks, the User Plane Function (UPF),  which  implements  equivalent functions to those present in 4G's Serving/Packet Data Network (PDN) Gateway (S/P-GW)\footnote{PDN Gateway (P-GW) and Serving Gateway (S-GW) can be implemented as one single entity called S/P-GW.},  anchors  the  IP  address  (as  the  4G P-GW module does) and can provide connectivity to a MEC instance through the 5G N6 reference point.
In addition, it can divert 
traffic with the ``UL CL'' (Uplink Classifier)  functionality  by  traffic matching filters.  In  this scenario,
UE traffic can be directed to different locations (i.e. Data Networks), for example 
when each traffic flow corresponds to a different application with particular requirements in terms of latency or other quality related parameters that  MEC applications must satisfy. Accordingly, our solution can be viewed as a practical implementation of the UL CL functionality that leverages SDN to provide advanced features such as dynamic deployment of new session anchors (UPFs) and application relocation from the core to the MEC or between different MEC locations, as shown in Fig. \ref{fig:5gintegration}. 

As another possible deployment scenario in 5G networks, ETSI  considers ``cloud wrapper MEC applications'' that provide Functions as a Service (FaaS) for Massive IoT devices. In this architecture,  
IoT 
traffic is sent to a cloud wrapper MEC application that implements the desired function, acts as a breakout to a cloud service provider or transfers the traffic to another MEC application with appropriate resources. Our solution avoids the need for specific cloud wrappers 
for different applications by providing a similar functionality with a generic implementation.

Summing up, current proposals to integrate MEC in 4G and 5G networks require complex setups whose components must be aware of the locations of the users and their own. Furthermore, these networks 
must be able to 
steer traffic to the  core or  any of the different MEC infrastructures, in order to achieve 
performance levels that depend on the proximity of the UE to the  applications running therein. 
In this paper we leverage virtualization and SDN to propose an architecture for that purpose, which we test on a functional 4G implementation. The destination of the traffic may change dynamically if the UE moves, the state of the network changes, the application is relocated, or the requirements of the UE vary. 
 
To achieve this goal, we create new instances of the S/P-GW, the termination point of UE traffic in 4G networks, in the desired location (core or MEC infrastructure).
Since the S/P-GW has a context state for each connection that must also be transferred,  we have also designed and evaluated a new mechanism to transfer and synchronize contexts between S/P-GW instances using SDN. This mechanism is completely transparent to the UEs and no protocol modification nor adaptation are necessary in  the S/P-GW. In addition, as shown in our Proof of Concept (PoC), this proposal can be used to easily integrate MEC infrastructures into 4G networks.

After discussing the background in Section \ref{sec:related-work} and defining the problem in Section \ref{sec:problem-statement}, our main contributions are:
 \begin{inparaenum}[(i)]
 \item  A feasible SDN solution for transparent dynamic steering of UE traffic to the network core or a MEC infrastructure, valid both for 4G and 5G networks, as well as 
 its implementation using standard technologies such as OpenFlow (OF) and Open vSwitch (OVS)~\cite{openvswitch}. The solution includes a method for replicating the state of the connections moved from existing S/P-GW entities to new ones using SDN (Section \ref{sec:proposed-solution}); 
 \item A comparative study of that method  versus possible alternatives for replicating that state, by analyzing the overhead in terms, for instance, of the time required for the replication and the amount of copied data (Section \ref{sec:analysis});
 \item An analysis of its impact in the data plane (e.g. analyzing the traffic processing overhead introduced in the switch)
    using \linelabel{line:onos_to_ovs}\hl{OVS} (Section \ref{sec:data_plane_analysis}).
    \end{inparaenum}

\section{Background and Related Work}
\label{sec:related-work}

MEC 
will  provide  service environments and cloud-computing capabilities at the edge of the mobile network in close proximity to mobile subscribers. One of its main goals is latency reduction, 
a key 5G requirement \cite{osseiran2014scenarios}.

MEC implementations involve several functions of the 3GPP specification of the 5G system architecture \cite{3gpp.23.501}. For example, 
concurrent access to different data networks (e.g. a central and an 
edge data network), and 
rules to send traffic to IP anchor points (UPFs, in 5G Standalone Architecture (SA)) connected to different Data Networks (DNs). The 5G Core Network can select a UPF and steer traffic to a DN based on UE subscription data, location, or
a request from the Application Function (AF). The AF can determine  traffic routing with traffic steering requests to the Session Management Function (SMF)
for single UEs or groups of UEs in particular areas. ``UL CL'' or IPv6 multi-homing functionalities can be used to divert traffic according to traffic matching filters, but the standard does not define mechanisms for traffic steering at the local access to the DN. Applications can also indicate that they can be relocated, or that they should be notified about certain traffic events or UE traffic in a particular area. Finally, Session and Service Continuity (SSC) features can support
UE and application mobility. 

Thus, the 3GPP specification 
provides a detailed list of the functions to implement MEC in 5G.
Many mechanisms, however, are 
only sketched. Furthermore, although the standard allows UEs selecting a SSC mode to ensure uninterrupted service, this may require active UE collaboration when the IP address changes or a new application server is selected. Dynamic relocation of applications would be impossible otherwise. Our proposal, which includes  a practical  implementation of a MEC architecture with SDN, allows  applications to be relocated dynamically, as it allows deploying new UPFs at different locations while keeping IP addresses and connections transparently to the UEs.

\subsection{MEC architectures for 4G networks}

As mentioned in Section \ref{sec:introduction}, integrating MEC architecture in 4G networks may also be useful, and it has been investigated by the ETSI and other research.
In \cite{sabella2016mobile},
at a conceptual level,
the authors suggest server/proxy Domain Name Server (DNS) configuration as a way of redirecting traffic to edge applications. They do not, therefore, consider the possibility  of SDN  as an enabling solution.
They also mention that MEC standardization is still  very immature and that it should be coordinated with ETSI
NFV standardization (the ETSI, indeed, is driving MEC standardization and implementation\footnote{\url{https://portal.etsi.org/tb.aspx?tbid=826&SubTB=826}}). 

Other researchers
have followed non-standard approaches based on SDN, NFV and cloud technologies for supporting mobile edge innovative services. For example, the mobile networks branch of Central Office Re-architected as Datacenter \cite{peterson2016central}, M-CORD \cite{mcord}, built an SDN scalable connectionless core using those technologies that reduces signaling overhead in Long Term Evolution (LTE) networks. However, this 
required modified Evolved Packet Core (EPC) entities.
In the Low Latency Multi-Access Edge Computing (LL-MEC)~\cite{nikaeinll}  open source MEC control plane 
of the Mosaic5G project \cite{mosaic5gllmec}, SDN switches carry out the user plane. 
Despite its similarities with our proposal, it is 
not 
interoperable 
with existing core network entities, since it replaces them completely.

\subsection{SDN and NFV to improve the network core}

Telecommunication operators have also shown interest in leveraging NFV and SDN
to 
enhance EPCs cost-effectively. The proof of concept in \cite{sdnEnabledProposal}, which ETSI ISG for NFV also reported in \cite{sdnEnabledResults}, presented an SGW and a PGW based on SDN and NFV 
that 

implemented the control plane 
in the SDN controller and the user plane in an SDN fabric. This approach preserved the GPRS Tunneling Protocol (GTP) protocols, GTP-Control Plane (GTP-C) and GTP-User Plane (GTP-U),
 from the perspective of the Mobility Management Entity (MME) and
 evolved Node B (eNB) entities, and  
demonstrated that SDN can manage operator network rules. Our approach also relies on SDN to handle user traffic but maintains the SGW and PGW entities unmodified.

\subsection{Dynamic traffic steering}
\label{dynster}

Regarding SDN control of packet cores,  
the Wireless \& Mobile Working Group of the Open Networking Foundation (ONF) pointed out the rigidness 
of current architectures \cite{sama2015software}. They defined Openflow extensions to handle PDN connections while keeping Quality of Service (QoS), accounting and online charging, and 
identified 
two extensions of the OpenFlow protocol to manage GTP-U packets, matchings and actions. We have not followed this approach because these extensions 
have not been implemented yet. Furthermore, we did not even consider the possibility of modifying the OpenFlow protocol. 
 We assume that any such solution should be available in the OpenFlow standard, SDN controller updates and in existing devices.

There exist different proposals for diverting traffic in 5G networks. The  virtualized  EPC gateway in \cite{heinonen2014dynamic}  switches dynamically the data plane for each user to a fast-path dedicated element near the eNB using an OpenFlow gateway with GTP extensions that keeps active sessions. Since the OpenFlow protocol does not support GTP, a workaround was needed to analyze
the content of the packets inside GTP-U tunnels. As described in \cite{nguyen2017sdn}, standardization groups are still discussing this aspect. In \cite{zabala2018towards}, the authors faced 
the same issue.
They applied an 
OVS patch that was not fully compatible with OpenFlow \cite{gtpuovspatch1}, so they programmed the flows by setting Secure Shell (SSH) connections with the devices. 

Therefore, some previous approaches that leverage SDN flexibility 
to divert individual user traffic to the network edge must  
analyze 
the content inside GTP-U tunnels, which is not directly available to SDN devices. 
 OVS patches that allow encapsulating and decapsulating the GTP-U protocol are a feasible workaround. We follow this same efficient approach.

\subsection{State replication}
Edge intelligence
is a key 5G enabler for low-latency  
applications such as self-driving vehicles, industrial automation, tele-medicine, and virtual and augmented reality \cite{simsek20165g}. Unlike previous approaches that simply considered traffic redirection to specific applications running on edge servers, we propose deploying the S/P-GW at the 
edge and reconfiguring the network dynamically to {\it move} some users or flows  to the new S/P-GW instance, thus reducing latency and core congestion. 
To achieve this, the state of the original S/P-GW must be replicated to the new S/P-GW. This must be 
transparent to 3GPP protocols, which,  as stated in \cite{nguyen2017sdn}, 
are not that flexible.
State replication of mobile network functions (mainly control plane entities, such as the MME) has already been tackled
with a typical three-layered architecture with a load balancer front-end, a layer of stateless workers and  a database layer with session context information \cite{prados2016latency, premsankar2015design}. In  \cite{banerjee2015scaling} the latter was removed by embedding the states in the workers. Obviously, this two-layer approach required proactive state synchronization among the workers, which was solved with a custom protocol over TCP. It still however introduced  state retrieval delay.

Generic state replication for mobile edge clouds has already been addressed.  In \cite{machen2018live}, the authors presented a 
framework for migrating active applications of virtual machines or containers. It involved transferring the entire RAM content of the source entity to the 
destination. 
The MEC proposal in \cite{cau2016efficient} described EPC-as-a-service stateful components, as well as a method to replicate the whole EPC state, i.e. the MME, the S/P-GW and the Home Subscriber Server (HSS) in different EPC instances.

Therefore, previous state replication schemes have imposed high latencies and overheads.
In this work, however, we propose a 
MEC solution that migrates S/P-GW modules to the edge without interrupting user sessions. It keeps the state of these modules by simply replicating 3GPP messages. Unlike previous approaches, it is generic, efficient and compatible with all communication and network management standards involved.

\section{Problem Statement}
\label{sec:problem-statement}

Centralized clouds  cannot satisfy stringent latency requirements, so 5G architectures will relocate some applications and network functions from the network core to the edge, that is,  closer to 
end users. The  3GPP  has considered MEC for this purpose, but there are still open challenges. One of the most relevant is traffic routing between mobile UEs and services 
without any ad-hoc MEC nor UE configuration, even in dynamic scenarios where a service that was initially instantiated in the  network core (or a remote cloud) may have to be relocated to the edge to reduce latency, without interrupting any ongoing user session.
Our main goal is solving this challenge, which we call ``transparent session and service continuity in dynamic MEC''.
Seamlessly relocating a service to the edge  also requires relocating the IP connection endpoint of the UE (e.g. the S/P-GW or UPF) to the same edge location.

To achieve this we propose to dynamically deploy a new virtualized S/P-GW module at the network edge 
and reprogram 
SDN entities automatically to divert any affected flows to the replicated S/P-GW.

This leverages  current network solutions, respects existing 3GPP and control management protocols without any modification and allows a rapid deployment  in a real network.

In case the S-GW and the P-GW are physically separated, it is just necessary to replicate the S-GW entity, in a similar way to the SGW-Local Break Out scenario presented in~\cite{giust2018mec}.

Two key issues must be addressed: 
{\it S/P-GW replication} and {\it 
traffic diverting at individual user level}.

Since we want to keep existing connections undisturbed, the initial S/P-GW should continue to serve all connections except those we \emph{move} to the edge. Next, we need to replicate the S/P-GW (so multiple S/P-GWs will be available in the network).
The original S/P-GW keeps a context for each user that 
identifies traffic exchange
with the eNB through the S1-U interface. 
Thus, somehow, the newly deployed S/P-GW must have access to those contexts  to  handle the corresponding packets. In other words, the context of the original S/P-GW must be replicated to the new one so that it can establish the tunnels and become fully operational.

\begin{figure}[ht!]
\centering
\includegraphics[width=0.99\columnwidth]{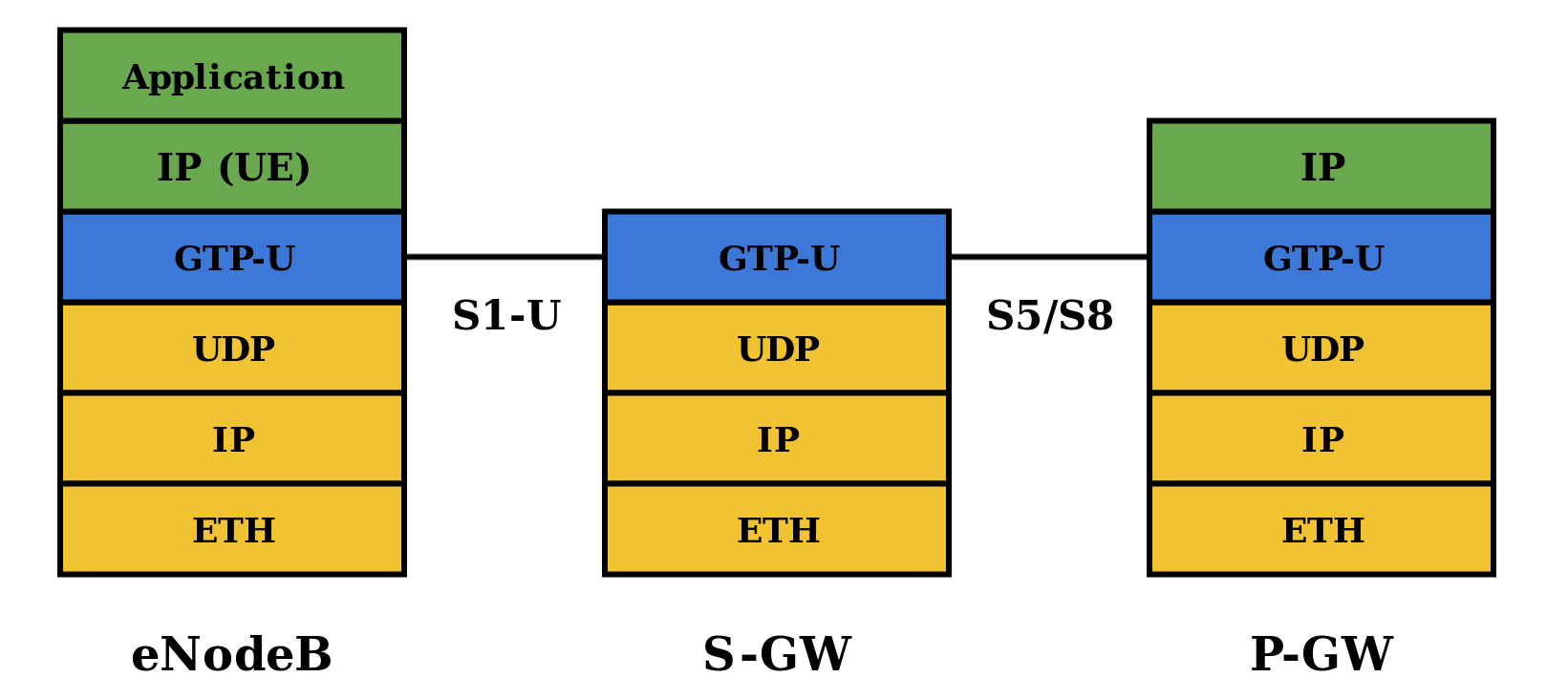}
\caption{GTP-U protocol stack in EPC interfaces}
\label{fig:user_plane_stack}
\end{figure}

It is also necessary to divert the packets
of specific users to the edge 
as soon as possible once the eNB has transmitted them. At this point, the packets are encapsulated with GTP headers whose outer IP headers correspond to the endpoints of the GTP-U tunnel: the eNodeB and the S/P-GW. The UEs, therefore, cannot be identified from 
outer IP or UDP headers, as they will be identical for all users connected to the same eNodeB. To divert traffic of specific UEs, the inner IP packets inside the GTP tunnels (i.e. the packets the UE sends) must be analyzed. Fig.~\ref{fig:user_plane_stack} shows the GTP-U protocol stack 
in the S1-U and S5/S8 interfaces of the EPC to illustrate this issue (outer packet layers  in yellow, GTP-U layer in blue and inner packet sent by the UE in green). 

\section{Proposed Solution}
\label{sec:proposed-solution}
As previously said, we propose an SDN-based architecture that  dynamically changes the endpoint of the IP tunnel for a particular UE. It allows deploying applications at the edge, and even relocating them when they previously run at other locations.

\begin{figure}[ht!]
\centering
\includegraphics[width=0.99\columnwidth]{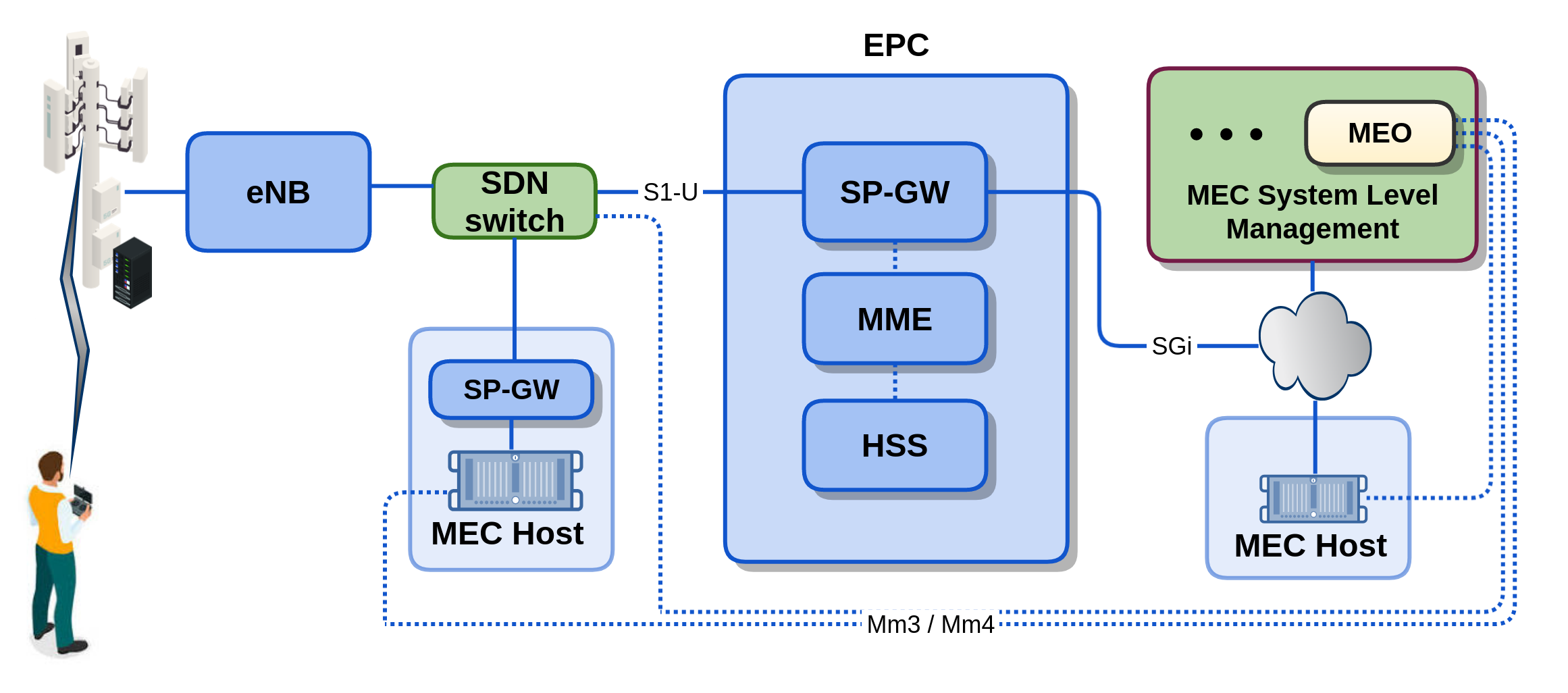}
\caption{Simplified view of the integration of our proposal in 3GPP 4G and ETSI MEC architectures}
\label{fig:sdn_ran_epc}
\end{figure}

First, our solution consists in including an SDN device in the architecture to handle the communications between each RAN and the EPC.
Fig.~\ref{fig:sdn_ran_epc} shows a simplified view in a 4G network. Initially, an SDN application running on top of the SDN controller manages the forwarding decisions of the SDN device with a simple reactive learning application, in the same way it manages the rest of the switches of the network.

Second, we propose an edge-assisted latency reduction framework  in two steps: (i) dynamic deployment of a new S/P-GW entity and (ii) an SDN application that replicates the state of the initial S/P-GW and configures the SDN fabric to divert traffic flows of specific UEs.

Note that our framework is fully compatible with the ETSI MEC reference architecture~\cite{etsi2019mecrefarch}. The proposed SDN device should be deployed in the S1 interface in a 4G network (Fig. \ref{fig:sdn_ran_epc}) or in the N3 interface in a 5G network (Fig. \ref{fig:5gintegration}). The SDN controller would be part of the MEC System Level Management, and, specifically, of the Multi-access Edge Orchestrator (MEO). The MEO might control the SDN switch either directly or as an AF (Application Function) in a 5G core.

The workflow of the solution is as follows:
\begin{enumerate}
\item The \hl{MEO} communicates with the Virtual Infrastructure Manager (VIM) through the Mm4 interface to launch a new S/P-GW Virtual Machine (VM) at the MEC host.
\item The \hl{MEO} communicates with the SDN application through the northbound interface of the SDN controller.
\item The SDN application then replicates the state of the original S/P-GW into the newly created one.
\item 
The SDN controller instructs the SDN device that communicates the eNodeB with the EPC to divert 
the user traffic that must be migrated to the new S/P-GW.
\end{enumerate}

The following subsections describe in detail 
this solution for S/P-GW state replication and traffic diverting at individual user level.

\subsection{S/P-GW replication}

Our solution  launches an S/P-GW VM in a server 
that is physically close to the eNodeB to which the UE is connected, or in a server that is close to the MEC application (for example, in the same MEC host where the application is running).
This S/P-GW must be fully operational, for which 
it must  access the state of the original S/P-GW. As discussed in Section \ref{sec:related-work}, the proposals in the literature employ fully stateless modules or stateful entities whose states must be synchronized, which 
requires custom modules 
and signaling protocols.
Conversely, we seek to address this step without modifying EPC entities and transparently to them.

Note that the S/P-GW is a reactive entity. 
We observed that the messages it receives from the MME determine completely its internal state, since the MME
instructs the S/P-GW to establish data path GTP-U tunnels (S1-U interface) with the eNodeBs. We propose to replicate the state of the original S/P-GW by sending  to the new S/P-GW the same GTP-C messages the MME has sent to the original S/P-GW.
The SDN application needs to store these original messages, and, when instructed to replicate the original state to a newly deployed S/P-GW, it will forward them accordingly. From this moment on, all GTP-C messages that the MME sends to the original S/P-GW (which the SDN application on the controller also receives), are sent to the new S/P-GWs as well.

This requires those GTP-C messages 
to be forwarded through the SDN controller,
for the SDN application to receive them. We have addressed this with a lightweight OVS switch in the MME virtual machine with  just one port in addition to the LOCAL port. This switch will treat all non-GTP-C packets transparently by sending the packets received at the LOCAL port to the second port and vice versa. A high-priority flow rule 
in the switch 
ensures that the GTP-C packets that the MME sends are forwarded through the SDN controller. A second high-priority flow rule, when traffic diverting is enabled, ensures that GTP-C packets generated by replicated S/P-GWs and addressed to the MME
are forwarded to the SDN application, which processes the packets as needed.
This guarantees that 
replicated S/P-GWs will be transparent to the original 
mobile network entities.
Note that the proposed SDN application is deployed on top of an SDN controller. Even though such controller is a  centralized entity in a logical network, it can be implemented as a physically distributed system to overcome scalability and availability issues, including mechanisms for failure handling, load balancing, etc.~\cite{kreutz2015software}. 

\subsection{Diverting  traffic at individual user level}

UE packets are encapsulated inside GTP-U tunnels. To redirect all user traffic to the replicated S/P-GW,  we could simply set a flow rule in the closest switch to the eNodeB to match all GTP-U traffic (i.e. UDP traffic for port 2152 \cite{etsi129281}) addressed  to the original S/P-GW and replace the destination IP and MAC addresses with those of the replicated S/P-GW.

However, we only want to divert the traffic of specific users and applications or even the traffic between specific users. Since the packets are GTP-U-encapsulated,  the GTP-U header carries user-specific information.
In order to differentiate users,  the switches must match the IP addresses of the inner IP header. 
However, the OpenFlow protocol \cite{openflow1_5_1} cannot match fields of layers above the transport protocol (UDP in the case of GTP-U) and, as discussed in  Section \ref{dynster}, extending the OpenFlow protocol to support GTP-U is not realistic 
because it would require modifying all the components that use that protocol, both switches and SDN controllers.
After analyzing several options, we found that OVS, the most extended OF-enabled switch, defines special  tunnel ports that allow encapsulating and decapsulating packets to access the inner UE IP packets. Even though the current version  
does not support GTP tunnel ports,
the patch in \cite{gtppatch1, gtppatch2} (also used in LL-MEC~\cite{nikaeinll}) extends OVS easily for that purpose.

This patch allowed us to create an OVS switch that is directly connected to the eNodeB, with a virtual GTP port 
that decapsulates inbound GTP packets and encapsulates outbound GTP packets. When a packet is encapsulated, the destination endpoint of the GTP tunnel of a flow can be set  with OVS  features that OpenFlow and most SDN controllers, such as Open Network Operating System (ONOS)~\cite{onos}, support in full.
Our solution uses three flow tables in the OVS switch that are managed by the SDN controller.

\begin{itemize}
\item Flow table 0: a classification table
\item Flow table 1: a GTP encapsulation table
\item Flow table 2: a forwarding table
\end{itemize}

The SDN controller  manages normal packet forwarding with a \emph{forwarding application}, and diverted traffic to replicated S/P-GWs with a \emph{diverting application}.
The forwarding application applies the following flow rules: the low-priority flow rule in table 0 matches any packet and its action is
checking table 2 for matches of that packet.
Table 2 is a regular forwarding table with a low-priority flow rule that sends any packet to the controller. Other medium-priority flow rules  in table 2 match source and destination MAC addresses to forward packets to the proper port. The SDN controller installs these medium-priority flow rules in the switches  as 
reactive learning 
applications. Whenever the controller receives a packet, it  learns the source MAC address and associates it to the corresponding switch input port. If the port associated with the destination MAC address is known, a medium-priority flow rule is installed in table 2 of the switch, which will directly forward the remaining packets of the flow at line rate. If the destination address is unknown, the packet will be flooded and no flow rule will be installed.

For example, to move \texttt{UE1} connected to eNodeB \texttt{ENB} from the original \texttt{S/P-GW1} to the new \texttt{S/P-GW2}, the diverting application will install the following flow rules in the closest GTP-enabled OVS switch to the eNodeB:

\begin{enumerate}
\item A medium-priority  rule in table 0 will identify GTP-U packets encapsulated by the OVS switch.
It will then replace the source IP addresses with that of the \texttt{ENB} and 
check for matches  in table 2,
for the packets to be forwarded as a normal ones.

\item A high-priority rule in table 0 will identify the packets that the OVS switch has decapsulated (i.e. those received at the \texttt{GTP} virtual port) and
check for matches in table 1,
 to  re-encapsulate the corresponding packets to their destination S/P-GW.

\item A low-priority  rule in table 1 will match any packet and send it to its input port (the \texttt{GTP} virtual port) for encapsulation. The rule will set the tunnel destination IP address to the \texttt{S/P-GW1} address, so  that the packets to be ignored will be encapsulated with a GTP-U header leading to the original S/P-GW,  with the GTP Tunnel Endpoint Identifier (TEID) of the original packet.

\item A high-priority rule in table 1 will match the packets sent by \texttt{UE1} and send them to their input port 
(the \texttt{GTP} virtual port) for encapsulation. This rule will set the destination IP address of the tunnel to the \texttt{S/P-GW2} address, so that UE1 packets will be encapsulated with a GTP-U header leading  to the replicated S/P-GW, with the GTP TEID of the original packet.

\item Finally,  a medium-priority rule in table 0 will identify GTP-U packets from  \texttt{ENB} for \texttt{S/P-GW1}. It will replace the destination IP and MAC addresses with those of the OVS switch, and send the packets to the \texttt{LOCAL} port for decapsulation. The network stack of the OVS switch will then receive the GTP-U packets, which  the GTP kernel module of the switch will decapsulate. The \texttt{GTP} virtual port of the OVS switch will then receive the decapsulated packets. The MAC and IP addresses of the GTP packets must be modified because these can only be decapsulated if the OVS switch is their final destination.
\end{enumerate}

Table \ref{tab:flowrulestable0} 
summarizes the flow rules 
installed 
in 
flow tables 0 and 1, after enabling traffic diverting for \texttt{UE1}. Note that forwarding table 2 is completely managed by the forwarding application, and is not affected by the diverting application.

\begin{table}
  \centering
  \caption{Flow rules installed in tables 0 and 1}
  \label{tab:flowrulestable0}
  \begin{tabu}{c>{\centering\arraybackslash}p{3.2cm}>{\centering\arraybackslash}p{3.2cm}}
  \toprule
     \multicolumn{1}{c}{\textbf{Priority}} & \multicolumn{1}{c}{\textbf{Match}} & \multicolumn{1}{c}{\textbf{Action}} \\ \midrule
     & \multicolumn{1}{c}{\textbf{Table 0}} &   \\ \midrule
    \texttt{HIGH} & \texttt{IN\_PORT=GTP} & \texttt{GOTO\_TABLE(1)} \\
    \texttt{MEDIUM} & \texttt{IPv4,NW\_SRC=ENB\_IP,} \texttt{NW\_DST=S/P-GW1\_IP,} \texttt{UDP,TP\_DST=2152} & \texttt{SET\_IP\_DST=OVS\_IP,} \texttt{SET\_ETH\_DST=OVS\_ETH,} \texttt{OUTPUT=LOCAL} \\ 
    \texttt{MEDIUM} & \texttt{IPv4,NW\_SRC=OVS\_IP,} \texttt{UDP,TP\_DST=2152} & \texttt{SET\_IP\_SRC=ENB\_IP,} \texttt{GOTO\_TABLE(2)} \\ 
    \texttt{LOW} & \texttt{ANY} & \texttt{GOTO\_TABLE(2)} \\ \midrule

     & \multicolumn{1}{c}{\textbf{Table 1}} &   \\ \midrule
    \texttt{HIGH} & \texttt{IPv4,NW\_SRC=UE1\_IP} & \texttt{SET\_TUN\_DST=} \texttt{S/P-GW2\_IP, IN\_PORT} \\ 
    \texttt{LOW} & \texttt{ANY} & \texttt{SET\_TUN\_DST=} \texttt{S/P-GW1\_IP, IN\_PORT} \\ \bottomrule
  \end{tabu}
\end{table}

\noindent

\begin{figure*}[ht!]
\centering
\includegraphics[width=0.99\textwidth]{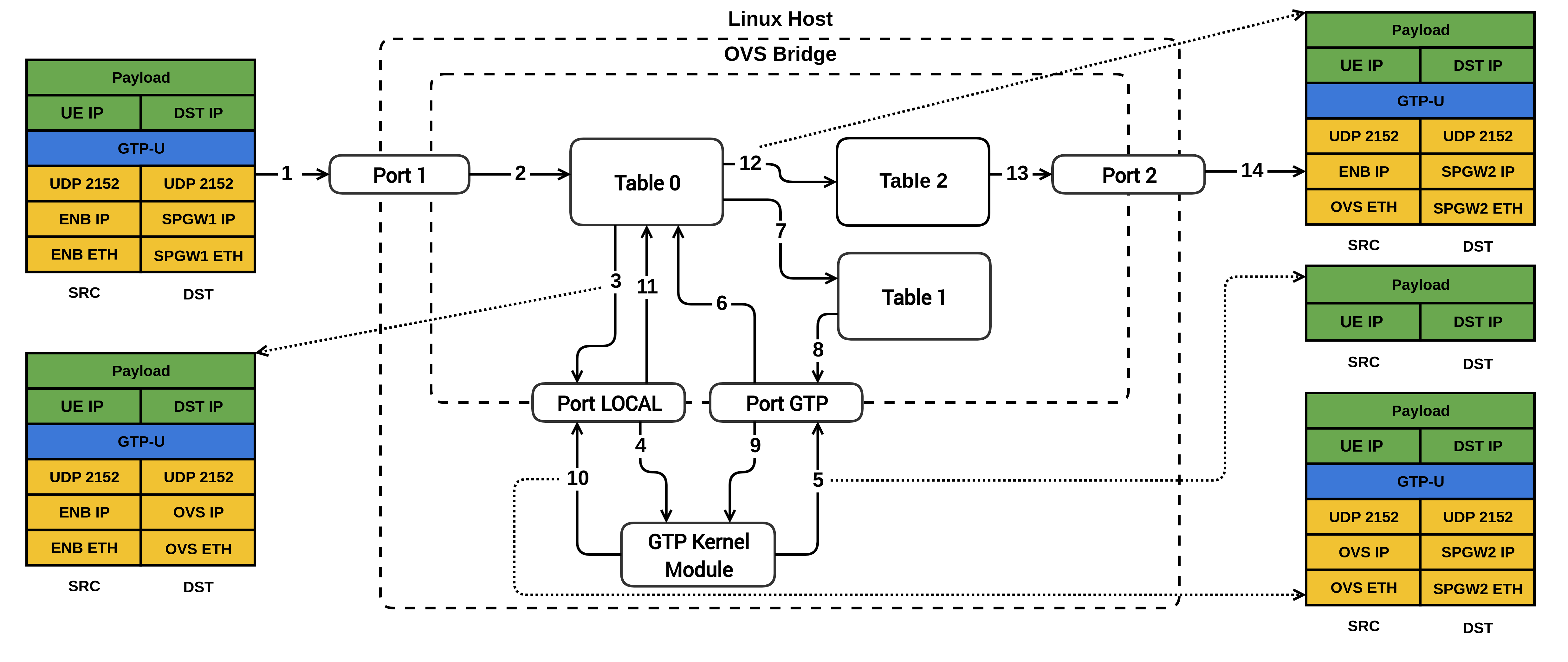}
\caption{Packet processing pipeline inside the GTP-enabled OVS}
\label{fig:block_diagram_data_plane}
\end{figure*}

Fig.~\ref{fig:block_diagram_data_plane} shows the  processing 
of a user plane packet in a GTP-enabled OVS that is close to the eNodeB. 
\begin{enumerate}
    \item The GTP-U packet is received at the physical port of the device (port 1 in the figure).
    \item The device checks for matches with flows in table 0.
    \item The packet matches a flow in table 0 that sets the destination IP and MAC addresses to those of the OVS as well as the packet output to the LOCAL port.
    \item The packet is then in the networking stack of the host, which forwards the packet to the GTP kernel module.
    \item This 
    module decapsulates the packet and forwards the inner packet to the bridge through the GTP port.
    \item The device checks for matches for the packet
    in table 0.
    \item The packet matches a flow in table 0, so the packet has to be
    checked against  the flow  entries in flow table 1.
    \item The packet matches a flow in table 1 that sets the destination IP address of the tunnel to the desired S/P-GW and the packet output  to the GTP port.
    \item The packet is forwarded from the GTP port to the GTP kernel module.
    \item The packet is then encapsulated with the desired S/P-GW as the destination GTP endpoint (with the same TEID as the original tunnel) and then inserted into the bridge through the LOCAL port.
    \item The device checks for matches in table 0 for the packet.
    \item The packet matches a flow in table 0 that sets the source IP and MAC addresses to those of the ENB, so the packet has to be
    checked in table  2.
    \item The packet matches a flow in table 2 that transfers the packet to the proper physical port.
    \item The GTP-U packet is forwarded through the physical port of the device (port 2 in the figure).
\end{enumerate}

\subsection{Characteristics of the solution}

We have implemented a fully operative solution in which:

\begin{itemize}
\item The locations of the traffic exit and entry points for the UE can be  dynamically modified thanks to the  replication of a fully operational stateful 3GPP-compliant S/P-GW, by replicating the GTP-C messages that the MME sends to the original S/P-GW through the S11 interface.
\item User plane traffic is diverted with existing technology to another S/P-GW at the network edge. No OpenFlow GTP extensions are necessary. An 
OVS patch encapsulates and decapsulates GTP packets transparently to OpenFlow.
\item Both the replication of a 3GPP-compliant S/P-GW and the dynamic 
user traffic diverting to this newly deployed entity are completely transparent to the network, and require no modifications to current entities.
As a result, our solution  is completely interoperable with 4G networks and can be quickly deployed.
\end{itemize}

Therefore, our solution is fully 3GPP-compliant and
is compatible with the OpenFlow protocol and with SDN controllers without any changes.
The only requirement is an existing OVS device patch for the OVS virtual tunnel ports to support the GTP-U protocol.

\linelabel{line:ssc_modes}\hl{Furthermore, our solution combines the benefits of SSC modes 1 and 3 defined by the 3GPP in TS 23.501\mbox{\cite{3gpp.23.501}}: it diverts the traffic to a new IP anchor point as in SSC mode 3 and, at the same time, it is able to preserve the IP address that was allocated to the UE as in SSC mode 1. Indeed, it overcomes the limitations of SSC mode 2 by following a \textit{make-before-break} approach, by replicating the state of the anchor point before redirecting UE traffic to the new IP anchor point. As a result, our proposal provides full session and service continuity in a completely transparent way to the UEs.}

\section{Analysis of the Impact of State Replication}
\label{sec:analysis}

As shown in  previous sections, our proposal can be easily integrated into existing 4G or new 5G MEC architectures. It can even handle application session continuity after handovers or MEC application relocations. In this section we analyze the impact of state replication.

\subsection{Initial considerations}
\label{sec:control_plane_analysis}

\begin{figure}[ht!]
\centering
\includegraphics[width=0.99\columnwidth]{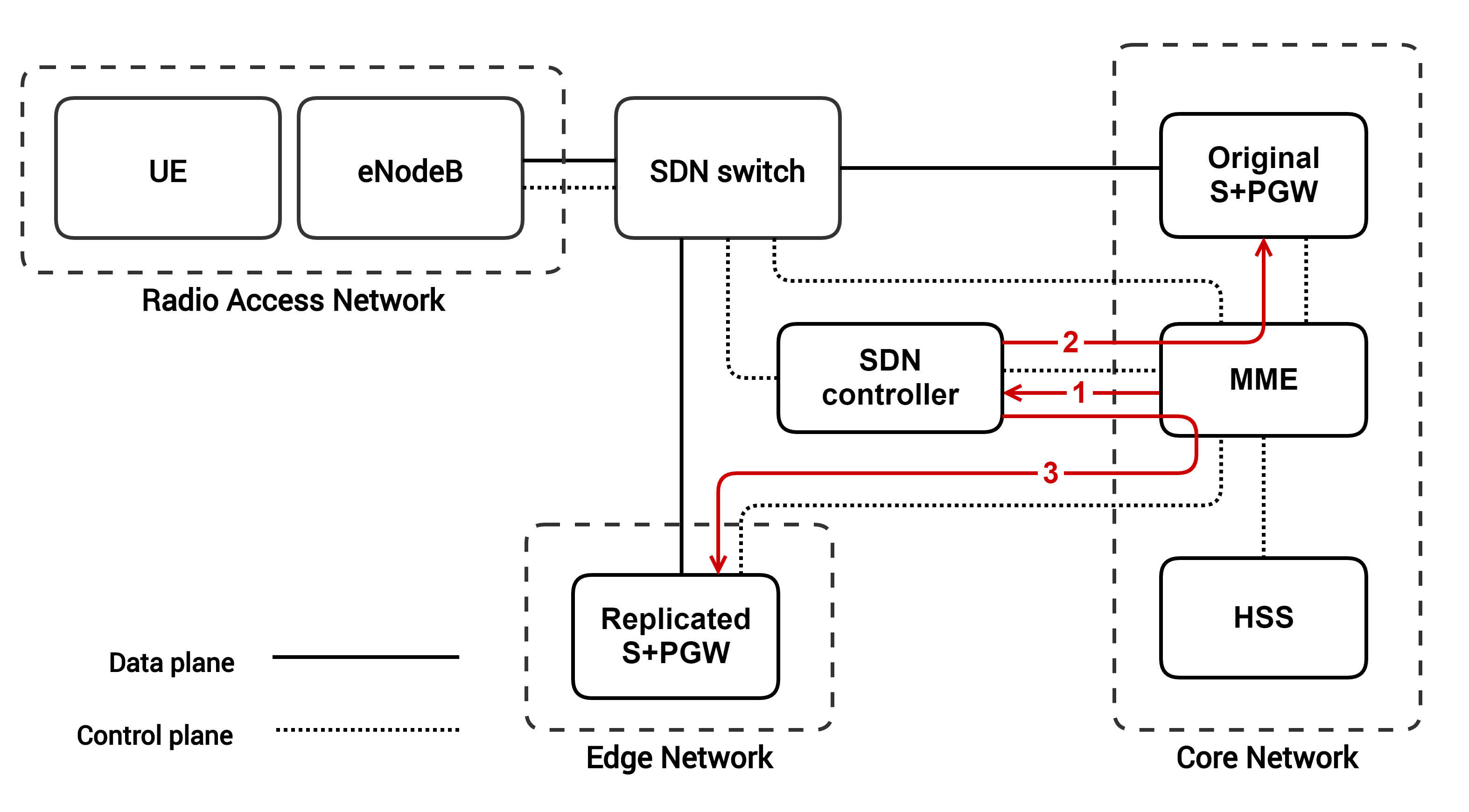}
\caption{\hl{Path for GTP-C messages from the MME to the S/P-GW, 4G case}}
\label{fig:extra_path_for_gtpc}
\end{figure}

The control plane handles the replication of the original S/P-GW context to the new S/P-GWs.
Our SDN application running on top of the SDN controller stores and forwards the GTP-C packets sent by the MME. When state replication is enabled, these stored packets are sent to the new S/P-GW. Fig.~\ref{fig:extra_path_for_gtpc} shows the path they follow. Control communications between the MME and the original S/P-GW include the GTP-C packets that the MME sends to the SDN controller in step 1. The SDN application stores received packets and sends them to the original S/P-GW through the MME switch in step 2. Finally, when state replication is triggered, the SDN application sends the packets to the replicated S/P-GW through the MME switch as shown in step 3.

First, we must consider 
MME GTP-C packet forwarding through the SDN application. It increases control traffic in the OpenFlow channel between the SDN application and the MME OVS switch, and
it clearly introduces some control communications overhead between the MME and the S/P-GW. 
The total latency of a control GTP-C packet between the MME and the S/P-GW will be higher than in normal  transmission. This extra latency 
is due to the transmission delay from the OVS switch to the controller; the SDN application processing delay; and the transmission delay from the controller to the OVS switch. The only processing required at the controller is to retrieve the destination IP address of the packet (i.e. the S/P-GW address) and to store the packet in the list of messages associated with that S/P-GW. Processing time at the SDN application is negligible (in the order of microseconds). The time required to send the packet to the controller corresponds to the transmission delay of an OpenFlow \texttt{PACKET\_IN} message containing the GTP-C packet. Analogously, the time required to transfer the packet from the controller to the OVS device is the time  to transmit an OpenFlow \texttt{PACKET\_OUT} message containing the GTP-C packet. The total measured  communication latency is in the order of 10 milliseconds, which is coherent with previous studies on   control plane latency of SDN-enabled switches \cite{he2015measuring}.

Even though a delay in the order of 10 milliseconds may look acceptable, we studied how to reduce it: Instead of sending each GTP-C packet to the SDN controller, storing it in the SDN application and then sending it back to the switch for normal forwarding to the S/P-GW, we considered the possibility of  sending the packet to the controller simultaneously and forwarding it directly to the S/P-GW  (e.g. including two output actions in one flow rule or using an OpenFlow ALL Group Action that sets outputs to the proper device port and the SDN controller). This approach would completely eliminate any extra GTP-C communications delay and allows asynchronous packet storage in the SDN controller.

The next relevant aspect is the {\it amount of information that must be stored in the system}. In our message replication approach the SDN controller is the  most involved entity,
since the SDN application it runs
must store the GTP-C messages that the MMEs send to the S/P-GWs.

The SDN application could store all the GTP-C packets that are received from each MME up to that moment. This simple solution is valid for a proof-of-concept, but it
does not scale well. From some moment on, it will be necessary to drop certain GTP-C packets stored in the SDN application  or transfer them to another storage system. However, this would affect the contexts of future S/P-GWs deployments. For example, it would not be possible to completely establish the GTP-U tunnel for a given user in a new S/P-GW, and therefore to move the user to a newly deployed S/P-GW.

To avoid this we propose the following improvements:
\begin{itemize}
\item Only messages associated with active MMEs will be stored. Accordingly, when a MME gets disconnected, all messages received from that MME will be removed.
\item Only messages associated with active S/P-GWs will be stored. Accordingly, when a S/P-GW gets disconnected, all received messages related to it
will be removed.
\end{itemize}

This solution does not incur in any loss of context information and it stores less information in the SDN application. Note that GTP-C packets are relatively small. For example, the two basic GTP-C messages that must be stored for each UE, which the MME sends when the users get connected to the network, are \emph{Create Session Request} and \emph{Modify Bearer Request}. They are less than 146 and 43 bytes long, respectively. This adds up to a total of 189 bytes of control information per UE.

Another
aspect is the \textit{time for state replication}. After launching the VM with the new S/P-GW, the SDN application sends to the new S/P-GW all the stored GTP-C messages associated with the original S/P-GW. This lasts for the time it takes the SDN application to send the packets via \texttt{PACKET\_OUT} OpenFlow messages.
Once the first packet is transmitted, the rest  
follow it in sequence. The  S/P-GW must have finished processing a given message before receiving the next. With OpenAir-CN, this processing
takes less than 10 milliseconds for \emph{Create Session Request}  and about 1 millisecond for \emph{Modify Bearer Request}.

Moving a user flow does not affect data plane traffic, since the flow is forwarded through the original S/P-GW until the 
new one with replicated state is ready. Hence, the initial control plane delay, as previously described, should be understood as the elapsed time since traffic diverting is triggered until the S/P-GW is operational.

Finally, the \textit{control information overhead} is the control information that is transmitted through the network during state replication.
Our solution must send all the packets that  the MME and the S/P-GW have previously exchanged. 
This is less than 200 bytes per each user that is connected to the S/P-GW.

Despite the improvements of the replication scheme in this section, it  still requires to store the GTP-C messages that are received at the original S/P-GW  in the SDN   application and then transmit them to the new S/P-GW. Therefore, we refer to this scheme as \textit{naïve message replication}. In the following section we present further improvements to
increase scalability.

\subsection{Selective Message Replication}
\label{sec:selective_message_replication}

The first additional improvement is only storing messages associated with active UEs. In other words, when a UE gets disconnected, all received messages corresponding to this UE will be removed, guaranteeing that the amount of stored information at any time is proportional to the number of active UEs. However, the transmission delay for all stored GTP-C packets may still be unacceptable for extremely dynamic scenarios in which many users are connected to an S/P-GW. To mitigate this, we propose only sending GTP-C packets associated with the UEs  to be moved. If other UEs must be moved later, sending their respective GTP-C packets
is enough. With this improvement, both the information overhead of the migration process and the state replication delay are independent of the number of stored GTP-C messages. Indeed, they are proportional to the number of migrated users.

We refer to the scheme that results from applying
these two additional improvements  to naïve message replication as \textit{selective message replication}. It scales much better but it requires further modifications to our proposal:
\begin{enumerate}
\item For each UE, after sending the \emph{Create Session Request}, the SDN application must extract the S/P-GW GTP TEIDs corresponding to the S11 and S1U interfaces as well as the IP address assigned to the  UE from the \emph{Create Session Response}. The SDN application must also store this information. Then, when it replicates the  original S/P-GW  UE state to a new S/P-GW instance, it has to modify the S11 GTP-C TEID  in the \emph{Modify Bearer Request} message for the new S/P-GW, so that the latter can accept the message. The overhead of this header modification is negligible.
\item
The data plane traffic that is forwarded to the
new S/P-GW must also be modified as follows:
First, the S1U GTP-U TEID in 
eNodeB packets for
the S/P-GW (uplink) must be modified with the information stored in the SDN application. Second,
the original  UE  address must be translated to the address that the
new S/P-GW assigns to the UE in the uplink
(vice versa in the downlink).
\end{enumerate}

The first modification can be easily implemented in the SDN application. It just requires a simple header change in control plane traffic. We can extend our data plane solution in a similar way to apply the
second modification, which involves header changes in data plane traffic at line rate. Specifically, we can leverage the GTP-enabled OVS device
to decapsulate the packets, perform the network address translations and re-encapsulate them by setting the proper TEID.

Note that these modifications require to store some information in the SDN application. In addition to the GTP-C packets associated with
active UEs, for each active UE that is migrated to the
new S/P-GW the application needs to store the S11 and S1U GTP TEIDs and the new IP address assigned by the new S/P-GW. Note however that this information is proportional to the number of migrated UEs, which will obviously not exceed the active UEs.

\subsection{Comparison with alternative solutions}
\label{sec:control_plane_comparative}

One of our main contributions is the SDN technique that replicates the state of a previous S/P-GW instance into a new one for seamlessly maintaining connections even when the endpoint of the GTP tunnel (the S/P-GW)
 changes. We must therefore compare
this technique
with alternative schemes in the literature that replicate the state of an application or network function.
We evaluate our naïve and selective message replication solutions versus stateless S/P-GWs \cite{prados2016latency, premsankar2015design}, RAM replication \cite{machen2018live} and a custom protocol \cite{banerjee2015scaling}:

\begin{itemize}
\item Naïve message replication of all GTP-C messages received by the original S/P-GW into a new S/P-GW.
\item Selective message replication of the GTP-C messages bound to the UEs to be moved to a new S/P-GW.
\item RAM replication (dump) from original to new S/P-GW.
\item Stateless S/P-GWs
whose states are centralized in a database.
\item Custom protocol to replicate the state of an S/P-GW.
\end{itemize}

\begin{table*}
  \centering
  \caption{Qualitative comparison of control plane replication approaches}
  \label{tab:comparative_qualitative}
    \begin{tabu}{lccccc}\toprule
    \multicolumn{1}{c}{\textbf{Approach}} & \multicolumn{1}{c}{\textbf{References}} & \multicolumn{1}{c}{\textbf{Modification of EPC entities}} & \multicolumn{1}{c}{\textbf{Selective replication}} & \multicolumn{1}{c}{\textbf{Downtime}} & \multicolumn{1}{c}{\textbf{Increased CP latency}}  \\
    \midrule
    Naïve message replication & Our proposal & No & No & No & No \\
    Selective message replication & Our proposal & No & Yes & No & No \\
    RAM replication & \cite{machen2018live} & No & No & Yes & No \\
    Stateless S/P-GWs & \cite{prados2016latency, premsankar2015design} & Yes & Yes & No & Yes \\
    Custom protocol & \cite{banerjee2015scaling} & Yes & Yes & No & No \\
    \bottomrule
  \end{tabu}
\end{table*}

First, we compared the schemes qualitatively by checking if they required modifications of any existing EPC entity, 
allowed for selective replication, implied service downtime or increased control plane (CP) latency.

Stateless S/P-GW modules with centralized state
must be modified to store  their states in a centralized database. This database increases the latency of control plane communications, but it allows for selective replication without service downtime. VM RAM replication  does not require any modification of previous EPC entities but
it does not allow
partial state replications. Also,
this approach needs to stop the services before dumping the RAM  \cite{machen2018live}, thus incurring in downtime. Finally, a custom optimized state replication protocol  for moving individual user flows to the edge would require modifying the S/P-GW to support the regeneration of specific GTP tunnels, to allow for selective replication without any downtime.

Our message replication proposals do not require any
S/P-GW modifications
and there is no downtime because the original S/P-GW does not participate in the replication and works uninterruptedly.
Besides, 
our proposals do not increase the latency of control plane communications since the packets can be sent asynchronously to the SDN application, as previously explained. Naïve replication involves complete state replication, but the selective version allows for fine-grained replication. 

Table~\ref{tab:comparative_qualitative} summarizes the qualitative comparison.

The solutions that need EPC modifications are not interoperable with existing deployments. Since interoperability is a basic requirement in our problem statement, we only carried out quantitative analyses of the three interoperable replication approaches,
in terms of amount of information that must be stored in the system, elapsed time of state replication and network overhead. We also studied the scalability of these three approaches
as the number of registered and moved user grows.

The testbed for these experiments had four VMs (1 vCPU, \SI{2}{\giga\byte} RAM, \SI{10}{\giga\byte} Disk) in an OpenStack cloud \linelabel{line:server2}\hl{running on an Intel\textregistered~Xeon\textregistered~CPU E5-2603 v3 @ 1.60GHz, interconnected through a network that was limited to} \SI{1}{\giga\bit\per\second}. Each VM executed one module: an ONOS SDN controller, one MME and two  OpenAir-CN S/P-GWs. We report average results of 10 independent executions.

\begin{figure}[ht!]
\centering
\includegraphics[width=0.99\columnwidth]{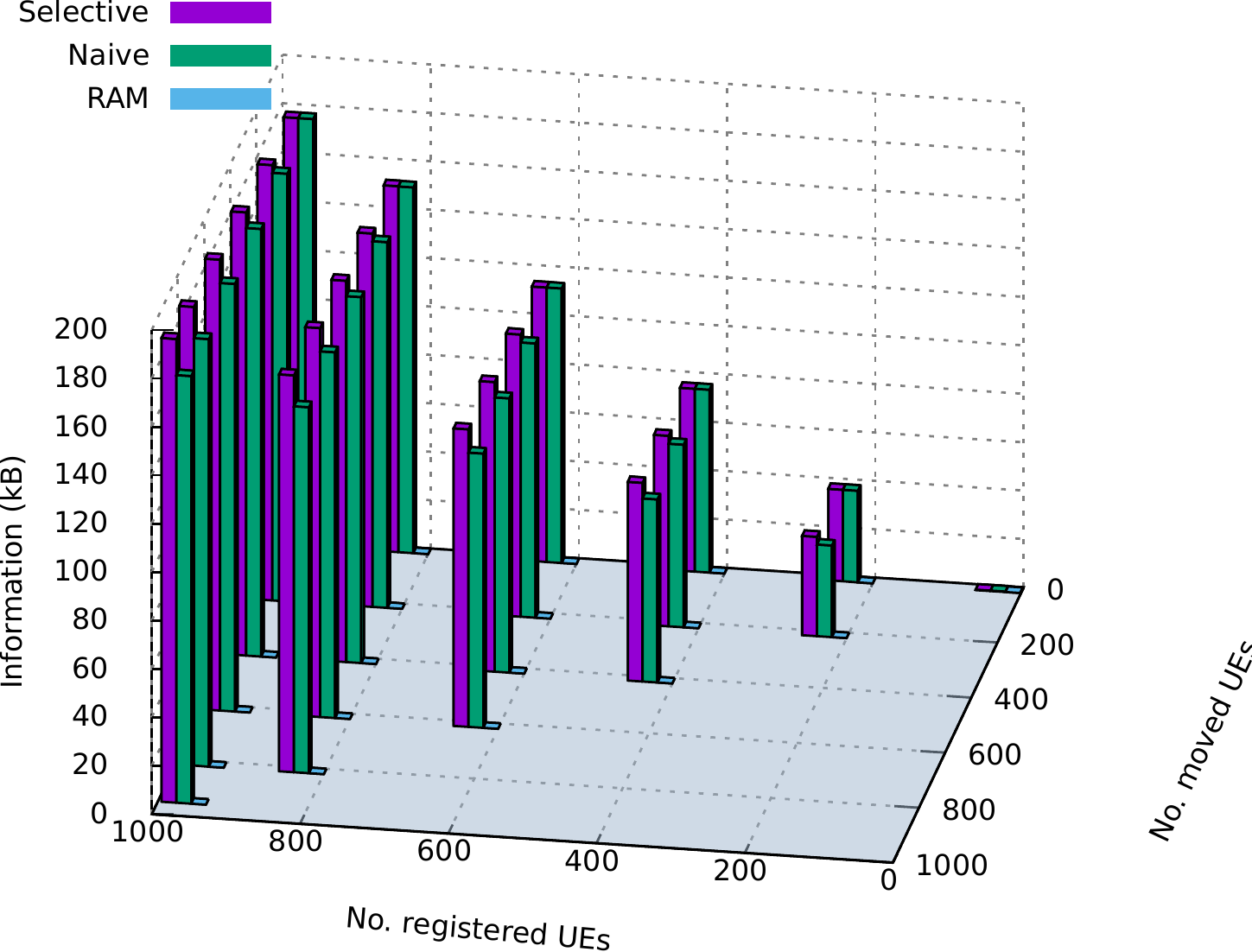}
\caption{Amount of information stored in the system}
\label{fig:control_plane_scalability_stored}
\end{figure}

Fig.~\ref{fig:control_plane_scalability_stored} shows the information the system must store as the number of UEs grows. Observe that the RAM replication solution does not store any information, since
the RAM content of the original S/P-GW is dumped to the new one. The information that our message replication proposals store, however, grows linearly with the number of users. In particular, naïve replication requires \SI{189}{bytes} per registered UE, whereas selective replication requires \SI{16}{bytes} per moved user plus \SI{189}{bytes} per registered UE. Note that these \SI{189}{bytes} per registered UE correspond to the \emph{Create Session Request} and \emph{Modify Bearer Request} GTP-C messages and \SI{16}{bytes} correspond to the S11 \& S1U GTP TEIDs (4 bytes each) and the IP addresses of the UE in the original and new S/P-GWs  (4 bytes each).

\begin{figure}[ht!]
\centering
\includegraphics[width=0.99\columnwidth]{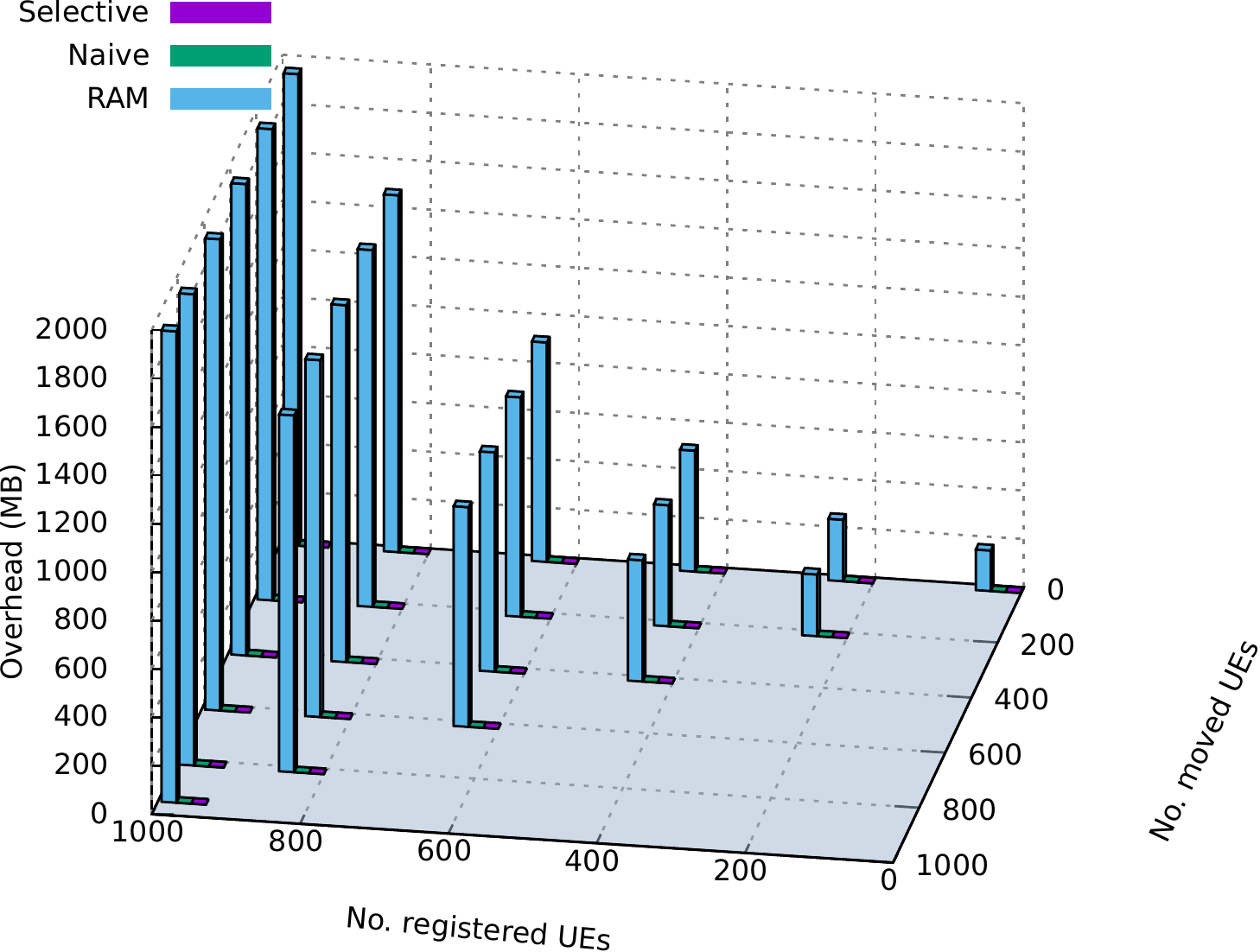}
\caption{Overhead introduced in the network}
\label{fig:control_plane_scalability_overhead}
\end{figure}

Fig.~\ref{fig:control_plane_scalability_overhead} shows the network overhead (control information) during a complete state replication.
It is considerably larger with RAM replication, ranging from \SI{160}{\mega\byte} for a single user to \SI{2}{\giga\byte} for 1,000 registered users. Moreover, with this approach the amount of stored information depends on the number of registered users. In addition, the overhead does not increase linearly  with the number of registered users, but quadratically. Conversely, the message replication approaches introduce minimum overhead: \SI{189}{bytes} per registered user with the naïve version and \SI{189}{bytes} per moved user with the selective version. Again, these \SI{189}{bytes} correspond to the \emph{Create Session Request} and \emph{Modify Bearer Request} GTP-C messages. For 1,000 registered and moved users, this is just \SI{189}{\kilo\byte}, three orders of magnitude less than with RAM replication, which requires almost \SI{2}{\giga\byte}. This highlights the potential of our message replication approaches, as they have minimum impact on the network, particularly the selective version.

\begin{figure}[ht!]
\centering
\includegraphics[width=0.99\columnwidth]{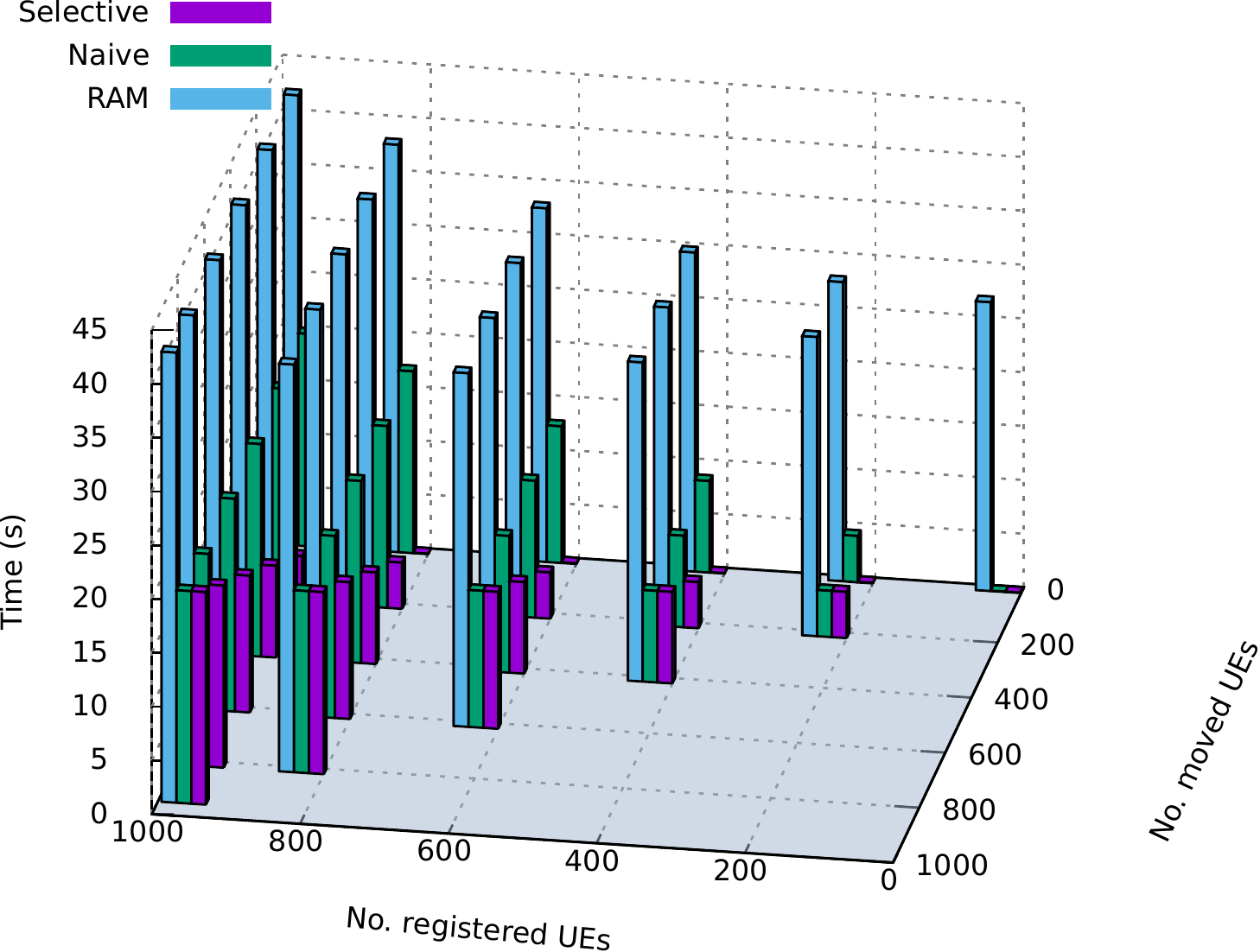}
\caption{Elapsed time for state replication}
\label{fig:control_plane_scalability_time}
\end{figure}

Fig.~\ref{fig:control_plane_scalability_time} shows the elapsed time of a complete state replication.   RAM replication is much slower than  our proposals for any number of registered or moved UEs.
Note again that RAM replication is not selective  when only certain registered UEs are moved. The difference is relevant when replicating the state for few registered users: for just one, over \SI{26}{seconds} with RAM replication  versus just \SI{50}{milliseconds} with our replication approaches. A breakdown of RAM replication times yields about 16.5 seconds to dump the RAM to disk, 1 second to transfer  \SI{160}{\mega\byte} of dumped RAM to the replicated S/P-GW and 8.5 seconds to restore the RAM from disk. We have also observed that the RAM dumping and restoration processes do not depend on the amount of RAM used, but on total VM RAM. 
The time to transfer the dumped RAM increases quadratically with the number of registered users, because it depends on the amount of transmitted information (as shown in Fig.~\ref{fig:control_plane_scalability_overhead}).
Conversely, the time for  naïve message replication  grows linearly with the number of registered users. With the selective version it only increases  with the number of moved users, allowing for faster replications. A breakdown of these times reveals
less than 1 millisecond to transmit a message and between 1 and 10 milliseconds for the  S/P-GW to process a received message and create the required context.

Summing up, the number of moved users has no effect on naïve message replication nor on  RAM replication, since they do not allow for selective replication. Consequently, their performance only depends on the number of registered users. Conversely, the performance of selective message replication  does vary with the number of moved users, since it allows for fine-grained replications.

\section{Impact  of User Traffic Diverting}
\label{sec:data_plane_analysis}

In this section we analyze our proposed solution for diverting traffic  at individual user level. We study its impact in the data plane 
to determine its minimum achievable delay and data overhead.
We also analyze its switch CPU load.

Our data plane solution requires an OVS switch 
with GTP-U capabilities close to an eNodeB. Its main CPU load  corresponds to   uplink packet decapsulation and re-encapsulation. Downlink packets are forwarded as normal traffic.

We measured 
its impact on data plane TCP overhead and packet delay.
We used as a baseline the scenario \textit{without GTP processing} in which GTP-U traffic is forwarded normally (without our solution). We compared this  reference with a \textit{GTP processing} scenario using our solution, whose performance is obviously lower. For the sake of clarity,
the measurements for the \textit{GTP processing} scenario took place after the flow rules were installed in the OVS switch and thus the network was in steady state.

\begin{figure}[ht!]
\centering
\includegraphics[height=2cm]{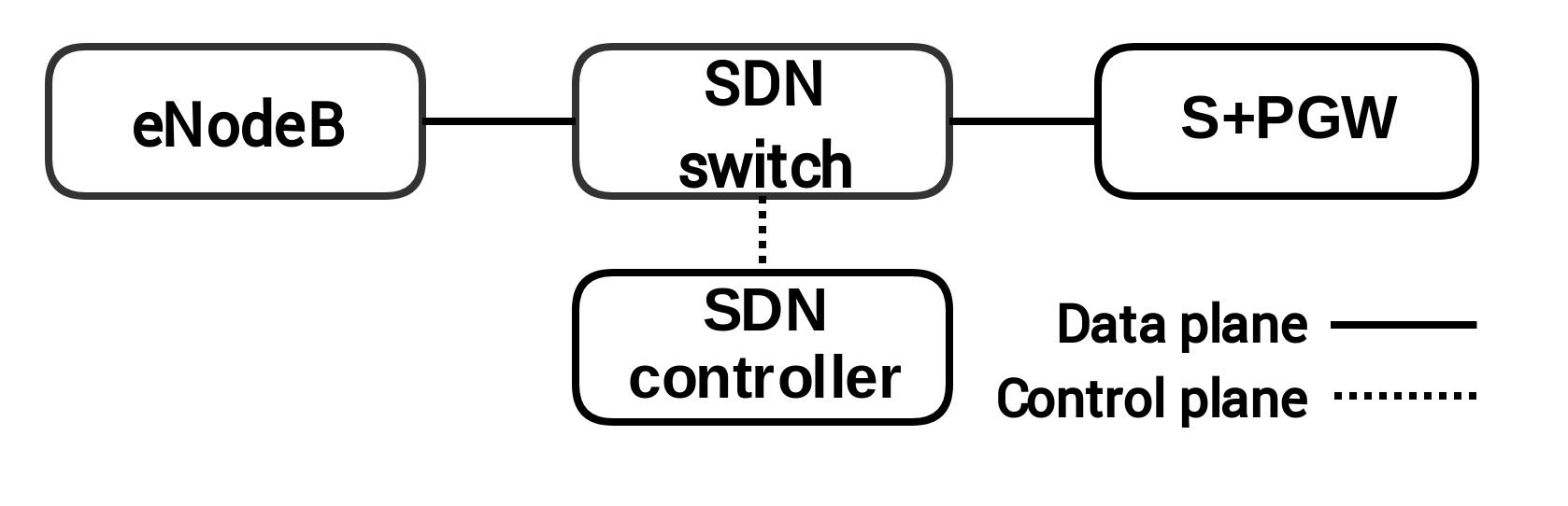}
\caption{Testbed for analyzing data plane overhead}
\label{fig:data_plane_scenario}
\end{figure}

The testbed for the experiments was a realistic setup with VirtualBox VMs (1 CPU and 2 GB RAM) \linelabel{line:server1}\hl{running on an Intel\textregistered~Core\texttrademark~i7 6700 CPU @ 3.4 GHz}. Fig.~\ref{fig:data_plane_scenario} depicts its architecture. Four VMs respectively ran the eNodeB, the S/P-GW, the ONOS SDN controller and the OVS switch. The  network interfaces of the first three  functions were connected to the OVS switch VM through different internal networks, forcing the traffic between the eNodeB and the S/P-GW to traverse the OVS switch.  We created a GTP tunnel between the eNodeB and the S/P-GW so that the traffic was GTP-encapsulated. Since GTP is a raw IP encapsulation protocol, we added static ARP entries to the eNodeB and the S/P-GW to enable connectivity between the GTP interfaces created at both endpoints. Inside the OVS switch VM, we created a GTP-enabled OVS switch as our solution mandates. In this switch we created a flow-based GTP tunnel port and we also included the interfaces that connected the VM with the eNodeB and the S/P-GW, resulting in the switch architecture in Fig.~\ref{fig:block_diagram_data_plane}. Finally, the SDN controller VM executed an instance of the ONOS SDN controller~\cite{onos}, with our SDN application running on top.

\begin{figure}[ht!]
\centering
\includegraphics[width=0.99\columnwidth]{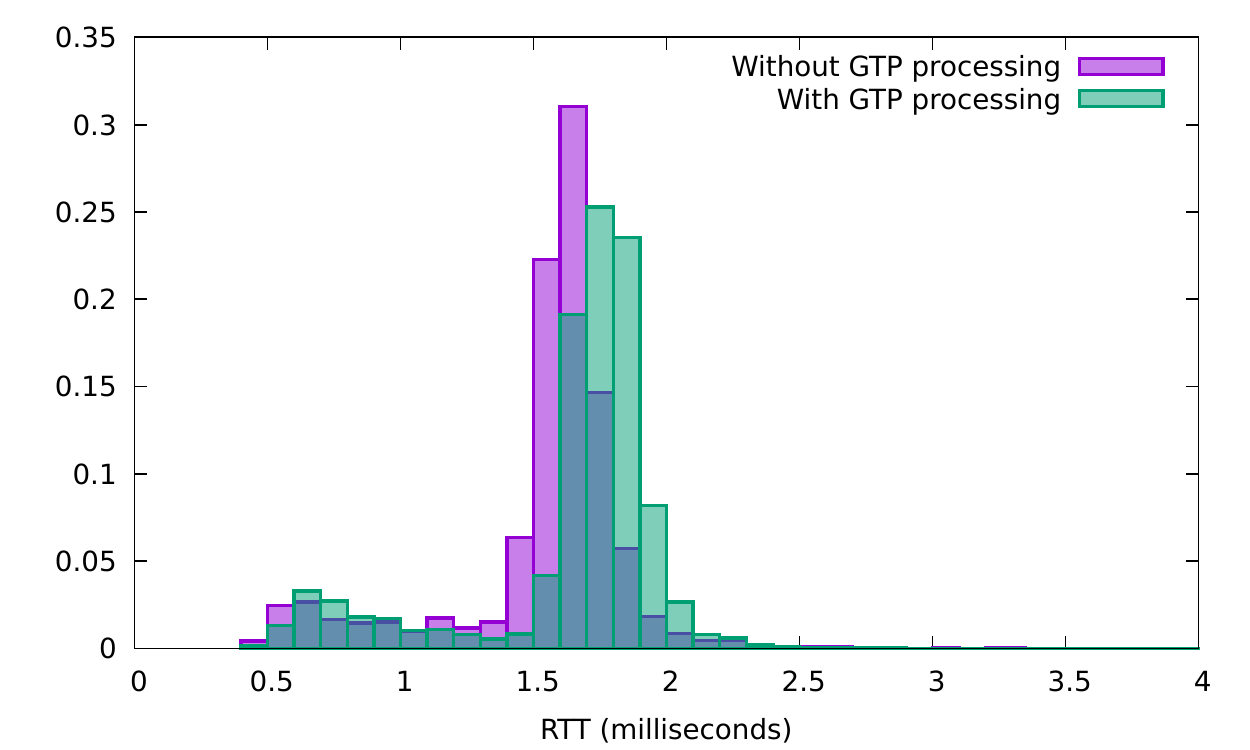}
\caption{Histogram of packet delay in the VirtualBox testbed}
\label{fig:rtt_histogram}
\end{figure}

The first experiment measured the communication latency between the eNodeB and the S/P-GW.
Fig.~\ref{fig:rtt_histogram} shows the histograms of the round-trip times (RTT) of ICMP ping packets.
The delay distributions were similar and approximately normal in both scenarios. The only difference were the medians (about \SI{1.7}{\milli\second} and \SI{1.6}{\milli\second} with and without GTP processing, respectively, indicating an RTT overhead of about \SI{0.1}{\milli\second} with our solution. The first conclusion from this result is that the  lower bound on achievable latency with our solution was \SI{0.1}{\milli\second}.

\begin{figure}[ht!]
\centering
\includegraphics[width=0.99\columnwidth]{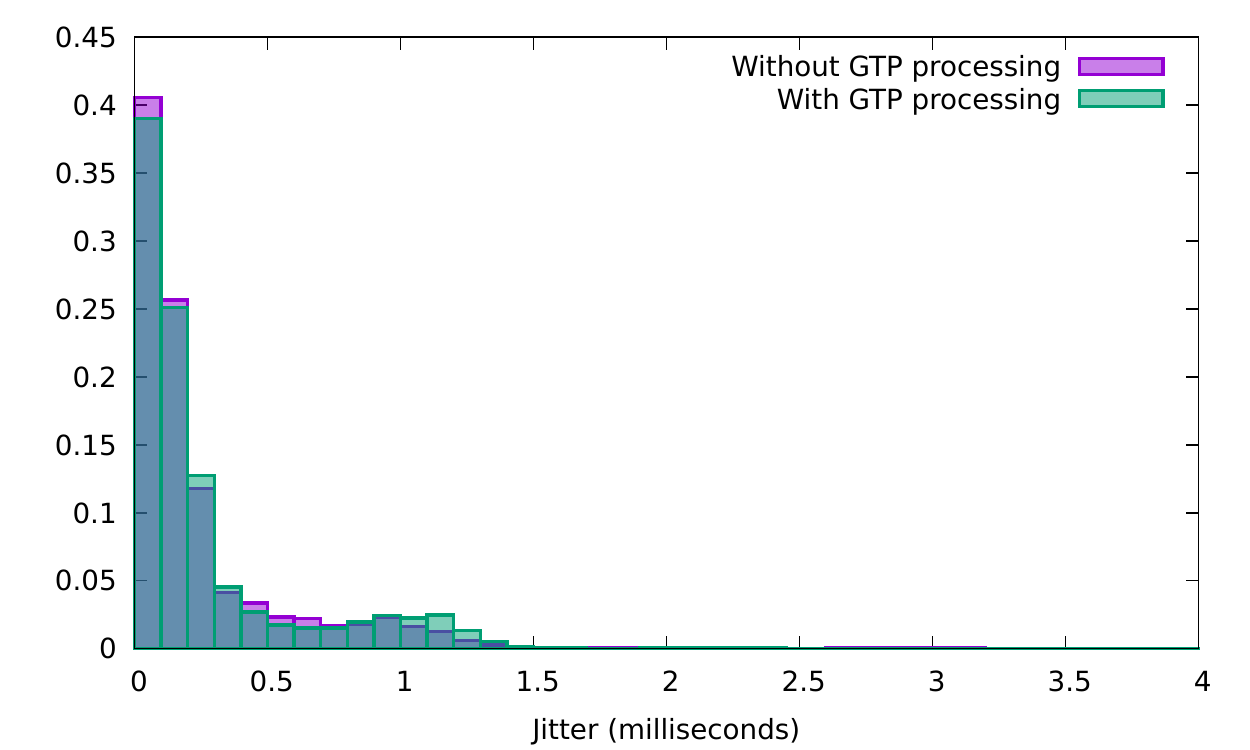}
\caption{\hl{Histogram of packet jitter in the VirtualBox testbed}}
\label{fig:jitter_histogram}
\end{figure}

\linelabel{line:jitter}\hl{In the following experiment we measured packet jitter between the eNodeB and the S/P-GW. Fig.~\mbox{\ref{fig:jitter_histogram}} depicts its histogram that was obtained as indicated in\mbox{\cite{rfc4689}}. As we can see both distributions are nearly identical. They follow a half-normal distribution with 75th percentile below {\SI{0.5}{\milli\second}}. Consequently, the impact of our solution on network jitter is negligible.}

\begin{figure}
  \centering 
  \subfloat[Achieved throughput]{
  \includegraphics[width=\columnwidth]{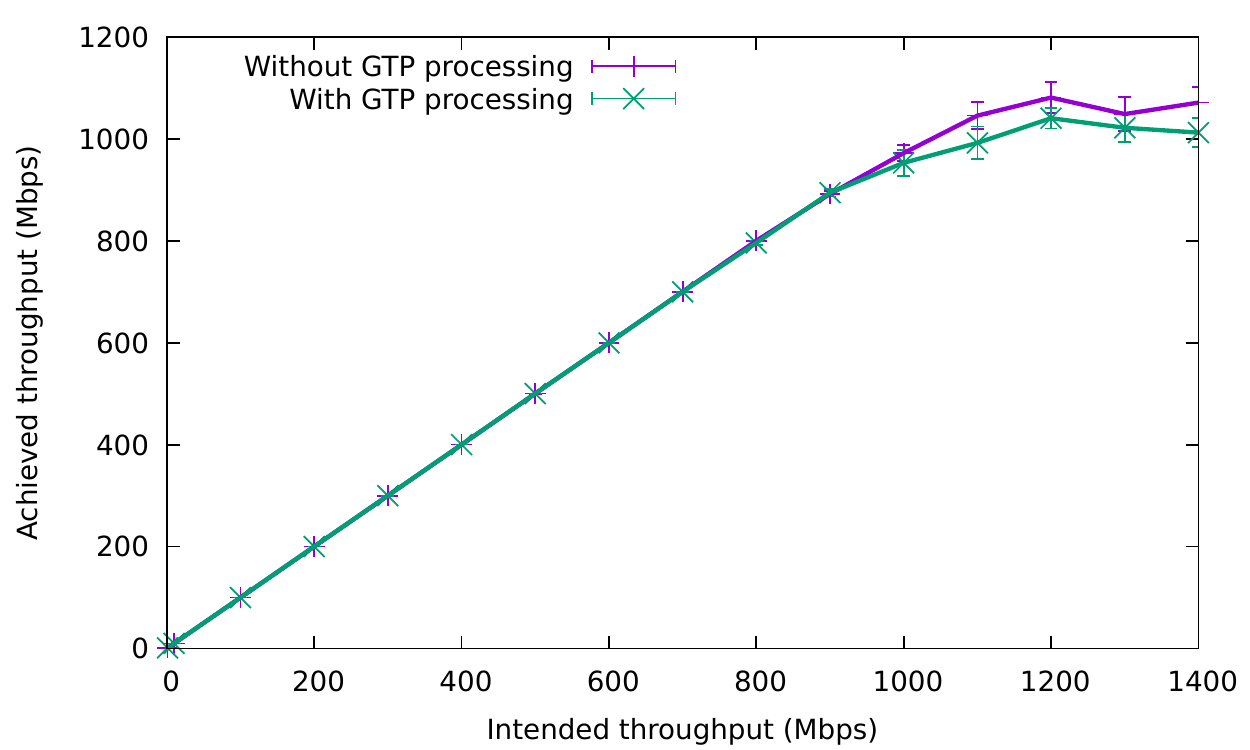}}\hfill
  \subfloat[CPU usage]{
  \includegraphics[width=\columnwidth]{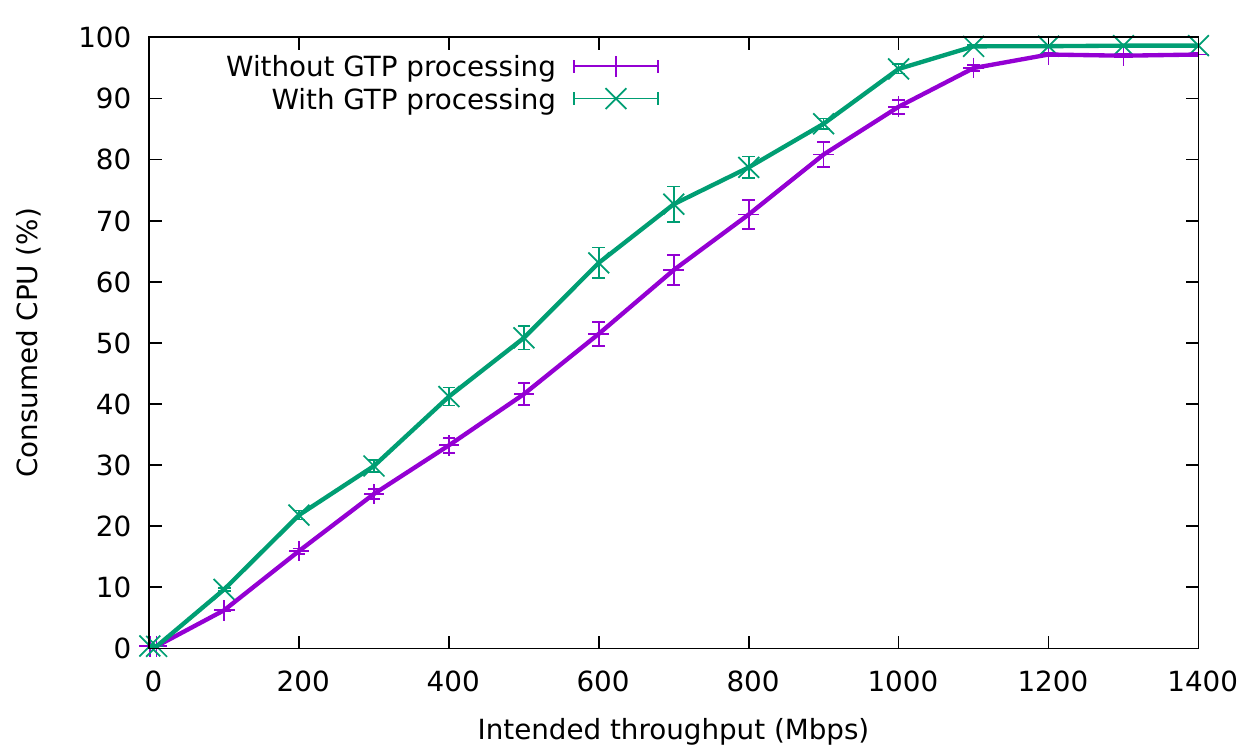}}\\
  \caption{Achieved TCP downlink throughput and CPU usage for different target throughputs}
  \label{fig:tcp_downlink_throughput_cpu}
\end{figure}

The next experiment measured the maximum TCP throughput  with our data plane solution. We  generated with \texttt{iperf3} an intended traffic rate  between the eNodeB and the S/P-GW  and measured how much of it was achievable. We also measured the CPU usage of the OVS switch VM to study its relationship with the overhead of our solution.
Fig.~\ref{fig:tcp_downlink_throughput_cpu} shows TCP downlink throughput and CPU usage for intended throughputs between 0 and \SI{1400}{\mega\bit\per\second}. Effective throughputs up to \SI{1000}{\mega\bit\per\second} were reached. From this point on, the downlink became saturated at about \SI{1050}{\mega\bit\per\second} without GTP processing  and at a slightly lower rate
with GTP processing. This small 
difference
in the downlink was also apparent in CPU usage: for the same target throughput, the scenario with GTP processing required up to \SI{10}{\percent} more CPU usage.

\begin{figure}
  \centering
  \includegraphics[width=\columnwidth]{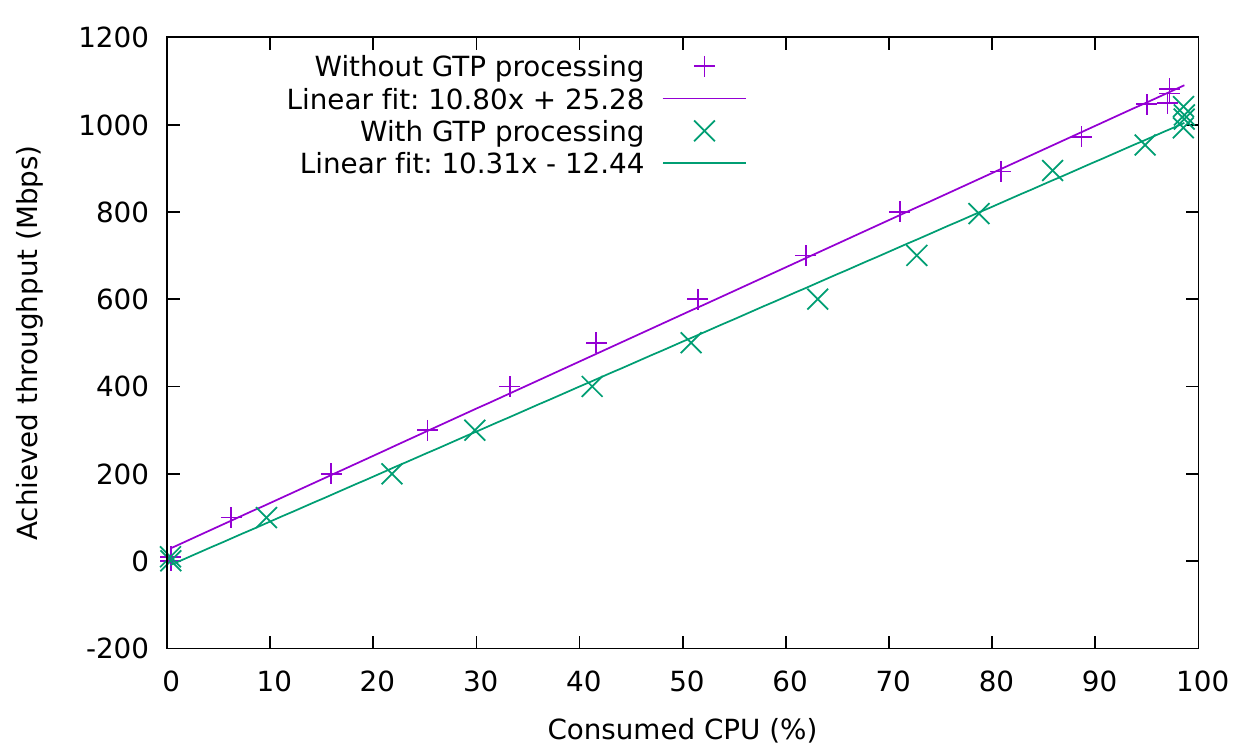}
  \caption{Relationship between achieved TCP downlink throughput and CPU usage}
  \label{fig:tcp_downlink_linear_fit}
\end{figure}

Even though the impact of our solution on downlink throughput was relatively low, we noticed an increase in  OVS switch CPU consumption. To understand the relationship between consumed CPU and achieved throughput, we estimated it as a linear regression as shown in Fig.~\ref{fig:tcp_downlink_linear_fit}. The correlation
was about 0.98 in both scenarios, confirming a strong linear relationship. The slope was practically the same except for minor differences due to the TCP ACKs in the uplink. The offset was noticeably higher  without GTP processing: for a given level of consumed CPU, about \SI{37}{\mega\bit\per\second}
extra throughput was achieved  without GTP processing. The offset was due to the processing and did not depend on traffic rate.

\begin{figure}
  \centering 
  \subfloat[Achieved throughput]{
  \includegraphics[width=\columnwidth]{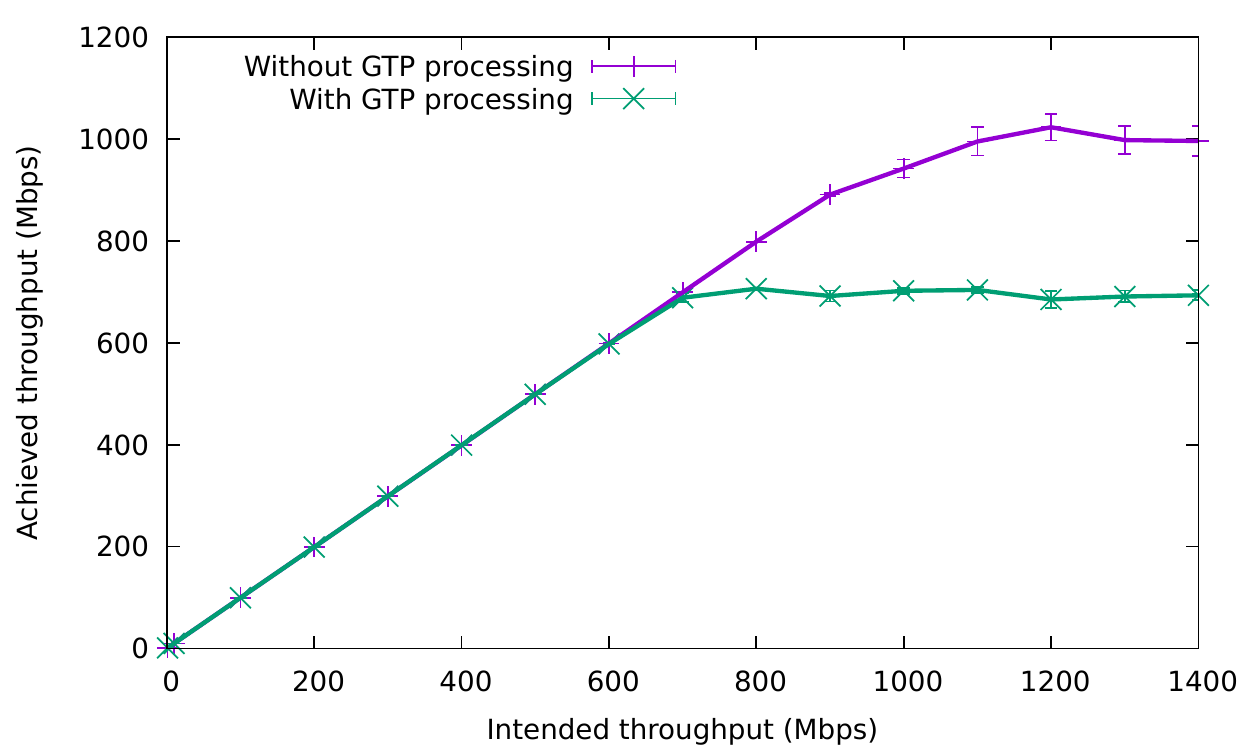}}\hfill
  \subfloat[CPU usage]{
  \includegraphics[width=\columnwidth]{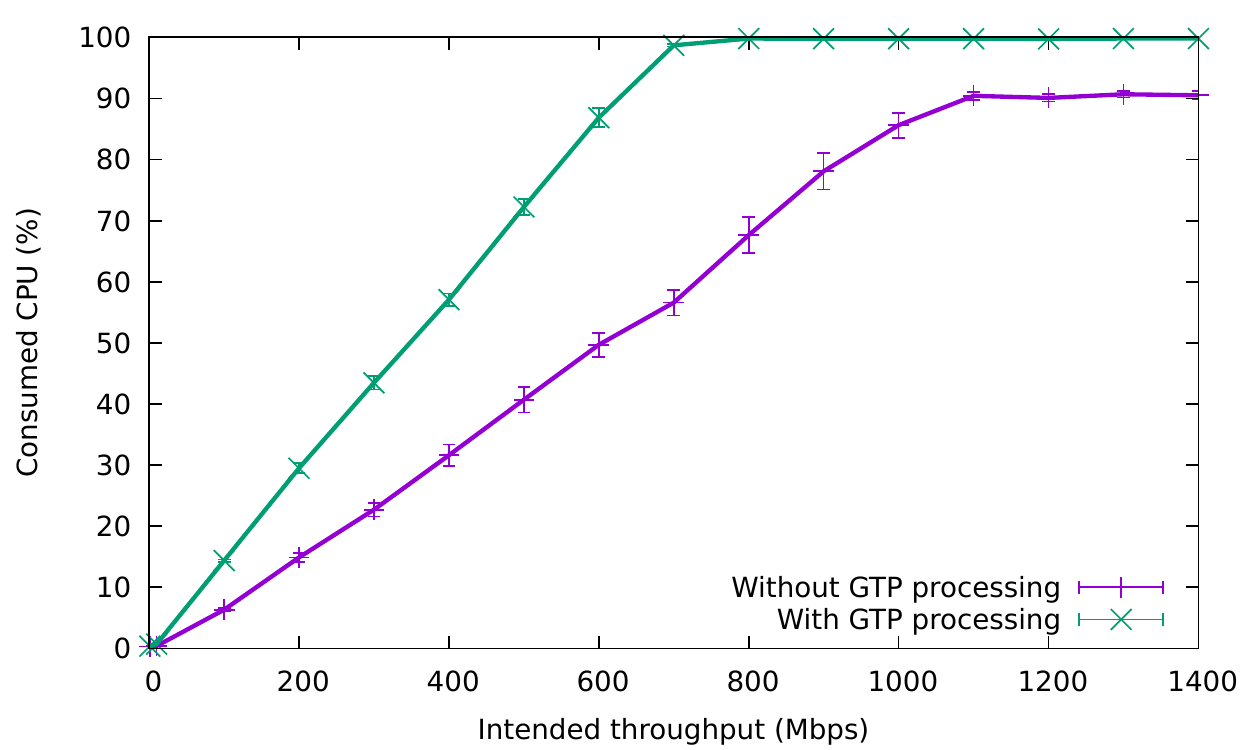}}\\
  \caption{Achieved TCP uplink throughput and CPU usage for varying target throughput}
  \label{fig:tcp_uplink_throughput_cpu}
\end{figure}

Fig.~\ref{fig:tcp_uplink_throughput_cpu} shows achieved TCP uplink throughput and CPU usage for intended throughputs between 0 and \SI{1400}{\mega\bit\per\second}. Maximum achieved throughputs were \SI{1050}{\mega\bit\per\second} in the baseline scenario and  \SI{700}{\mega\bit\per\second} with GTP processing. A different CPU usage behavior was also observed. With GTP processing it grew rapidly to \SI{100}{\percent}  at about \SI{700}{\mega\bit\per\second}. GTP processing had a noticeable impact on maximum achievable uplink
throughput. Note however that this was only due to CPU processing in the OVS switch, which  was implemented in our experiments in a VirtualBox VM. In  a real deployment with a physical SDN switch or a specialized computer the behavior would be similar yet at much higher rates.

\begin{figure}
  \centering
  \includegraphics[width=\columnwidth]{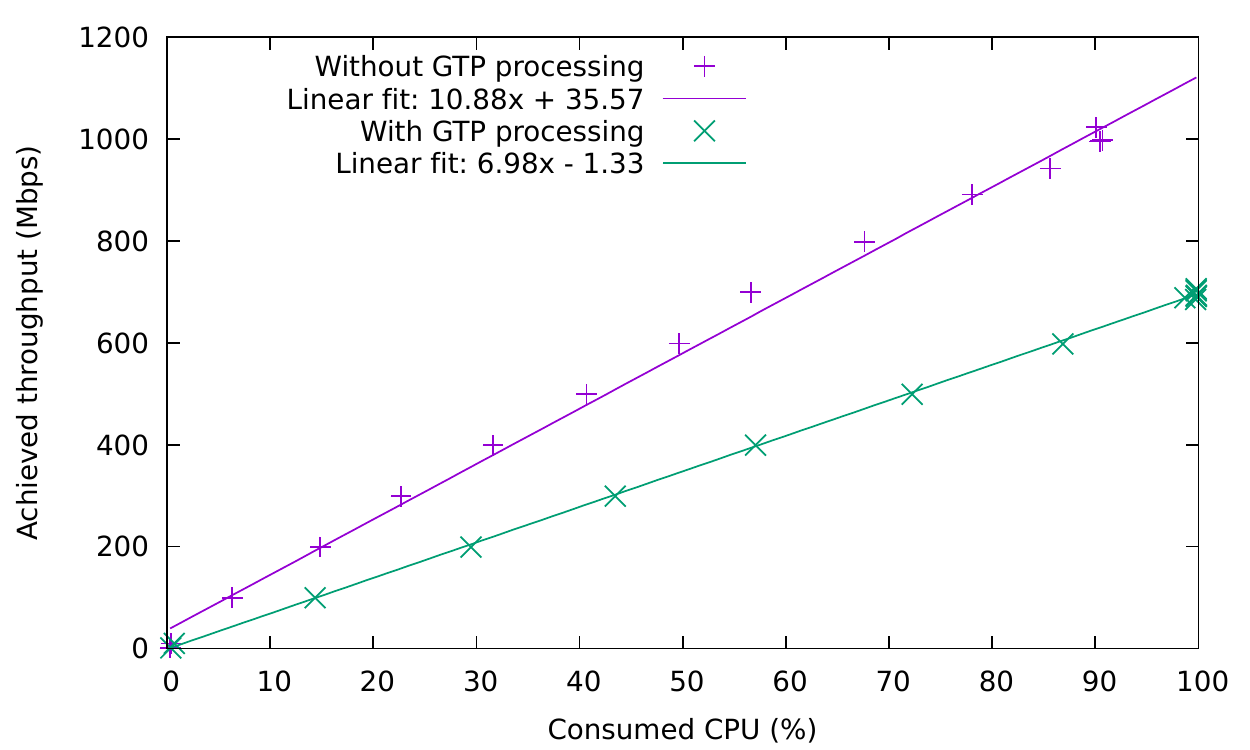}
  \caption{Relationship between achieved TCP uplink throughput and CPU usage}
  \label{fig:tcp_uplink_linear_fit}
\end{figure}

We also estimated the
linear regression between CPU usage and maximum achievable throughput.
Again, as shown in Fig.~\ref{fig:tcp_uplink_linear_fit}, the relationship was strongly linear, with correlations above 0.99 for both scenarios. The linear fit without GTP processing was similar to the downlink case, as expected. The
penalty
due to GTP processing was about \SI{37}{\mega\bit\per\second}, but
now the slopes were clearly different. A \SI{1}{\percent} increase in CPU load yielded a \SI{10.88}{\mega\bit\per\second} increase in achievable throughput in the baseline scenario, whereas
only a \SI{6.98}{\mega\bit\per\second} increase with GTP processing. This is because the extra
CPU usage of OVS GTP processing
is mainly due to the uplink. This CPU overhead has a linear relationship with traffic rate, so that just  $\frac{6.98}{10.88}\times 100 = \SI{64}{\percent}$ of the target throughput is achievable for any given CPU usage. The overhead is needed for flow-specific traffic diverting, which requires packet decapsulation and re-encapsulation in the OVS. Moreover, even in this simple setup using VirtualBox VMs on commodity hardware, over \SI{600}{\mega\bit\per\second} TCP throughput could be achieved in the uplink channel, which is acceptable in practice \linelabel{line:throughput_ok_cloud_gaming}\hl{not only for cloud gaming, which could need up to {\SI{10}{\mega\bit\per\second}}\mbox{\cite{cloud_gaming_bandwidth}}, but} even for bandwidth-demanding MEC applications such as AR, which may require up to \SI{30}{\mega\bit\per\second}~\cite{braud2017future}.

\begin{figure}
  \centering 
  \subfloat[Downlink]{
  \includegraphics[width=\columnwidth]{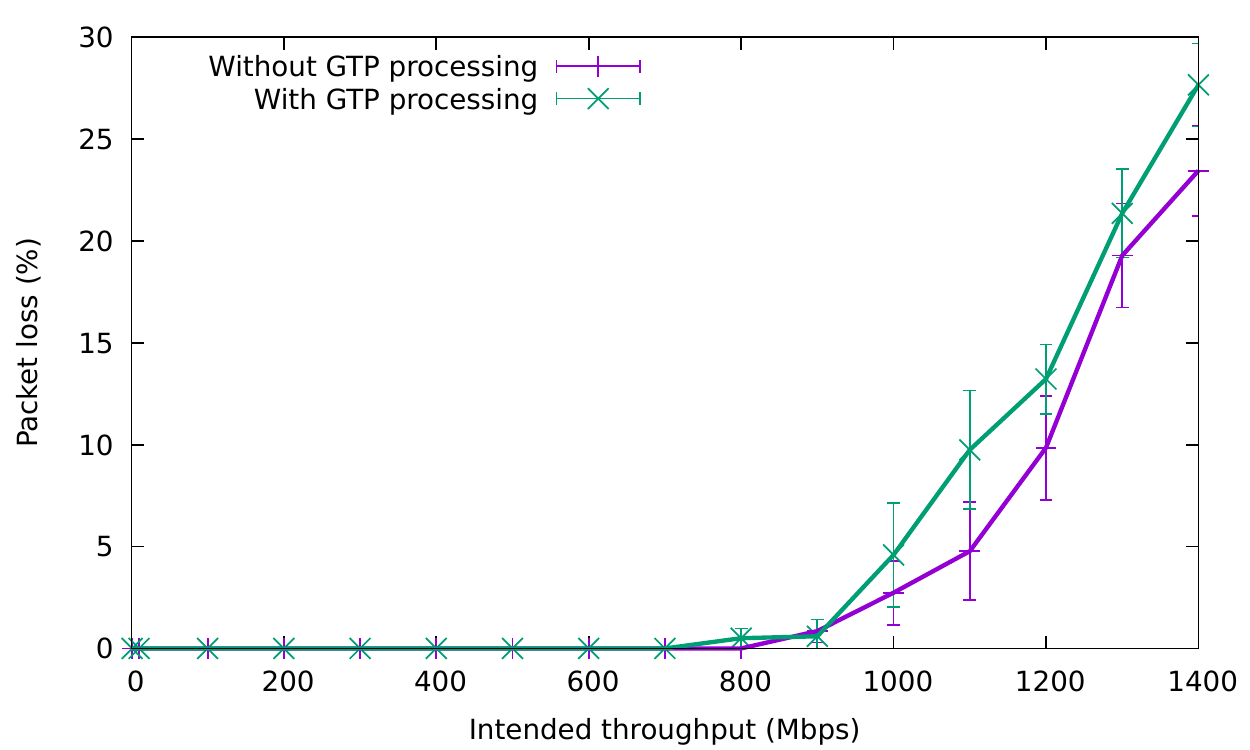}}\hfill
  \subfloat[Uplink]{
  \includegraphics[width=\columnwidth]{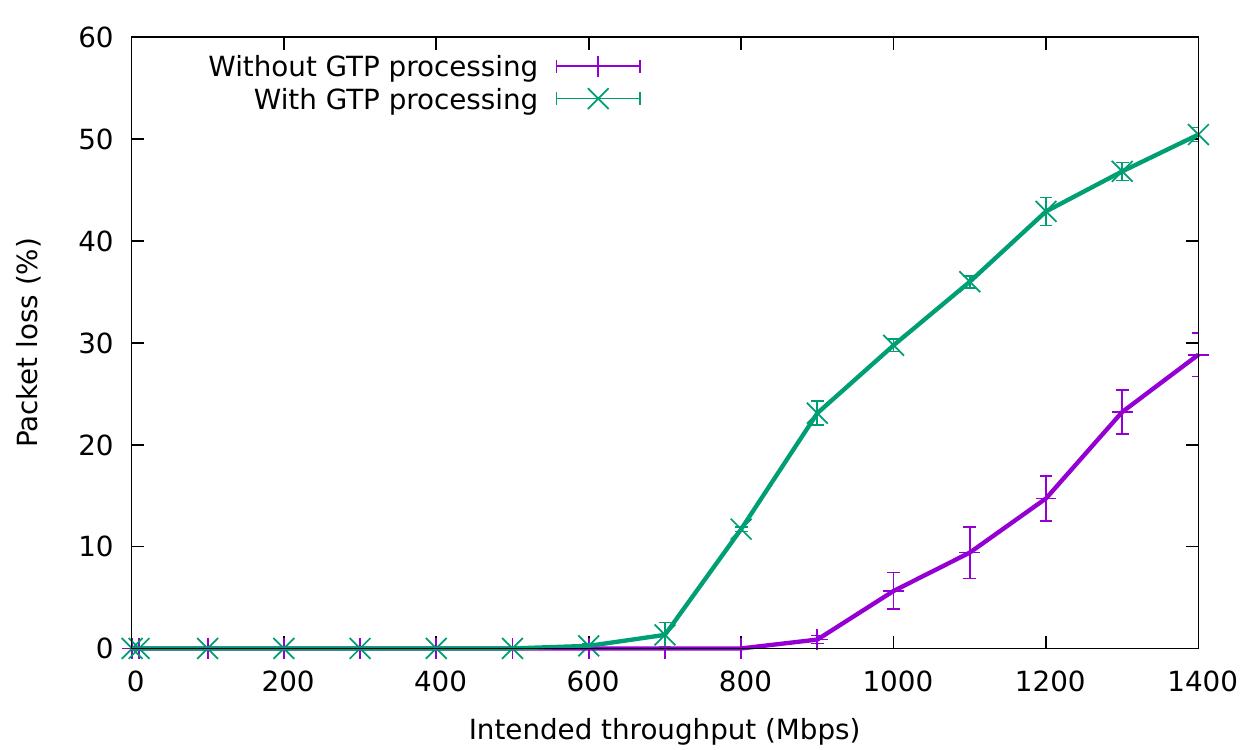}}\\
  \caption{\hl{Packet loss for different target throughputs}}
  \label{fig:tcp_throughput_loss}
\end{figure}

\linelabel{line:packet_loss}\hl{Our final experiment evaluated the packet loss between the eNodeB and the S/P-GW for different target throughputs. Fig.~\mbox{\ref{fig:tcp_throughput_loss}} shows the results of the experiment both in downlink and uplink direction. We can see that no packet loss was experienced below {\SI{600}{\mega\bit\per\second}} in any direction, and that it was still less that {\SI{1}{\percent}} in the uplink for any rate below {\SI{900}{\mega\bit\per\second}}. These values are in line with the saturation that was observed in  achieved throughput and thus higher intended throughputs introduce non-negligible packet loss.} \linelabel{line:packet_loss_ok_cloud_gaming}\hl{This analysis  validates that our solution can handle {\SI{600}{\mega\bit\per\second}} without packet loss, hence satisfying the requirement of sub-{\SI{1}{\percent}} packet loss for cloud gaming applications\mbox{\cite{itutg1032, lantecy_packet_loss}}}.

\section{Conclusions}
\label{sec:conclusions}

This paper presents an approach to
satisfy the latency requirements of services implemented in edge computing nodes. It allows
relocating applications from third party clouds or  the network core  to the edge.
Therefore, with this approach it is not necessary to deploy the applications at edge locations permanently.  It is also valid for transferring applications between MEC locations when the users move between them.

The approach
relies on SDN technology to redirect user traffic to  new traffic endpoints (SP-G/W or UPF modules) without interrupting ongoing user sessions. 
For that purpose,
the new modules must maintain the contexts of the previous instance. We have proposed a novel method
to replicate the state of the previous S/P-GW.
Essentially, the control messages
for that previous S/P-GW are captured and resent to the new S/P-GW  instance to replicate the original context.  

We have analyzed the overhead of this state replication solution
as well as the  SDN switch load due to the traffic that has to be redirected to the new endpoint. We compared our state replication method  with other possible  alternatives. The results indicate that our approach is highly competitive
even for large user populations.

The experiments also show that
a low-end VM with an OVS switch can redirect substantial traffic without significant delay or throughput impairment.

As future work we plan to  study the  case of mobile users accessing network-side applications with varying requirements. We are currently modeling user mobility and behavior and, from the results in this paper, quantifying operation parameters. Then we will study the scalability of inter-MEC handover with our solution.

\section*{Acknowledgements}

This work  has been supported by ``la Caixa'' Foundation (ID 100010434) fellowship LCF/BQ/ES18/11670020, MINECO grant TEC2016-76465-C2-2-R, and Xunta de Galicia grant GRC 2018/053, Spain.

\bibliographystyle{IEEEtran}
\bibliography{IEEEfull,main.bib}

\begin{IEEEbiography} [{\includegraphics[width=1in,height=1.25in,clip, keepaspectratio]{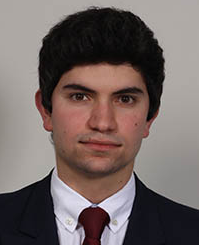}}]{Pablo Fondo-Ferreiro} received the Bachelor's Degree in Telecommunication Engineering in 2016 from University of Vigo, receiving the award for the best academic record. In 2018 received the Master's Degree in Telecommunication Engineering from the same university, holding the best academic record. In 2016 he received a collaboration grant from the Spanish Ministry of Education to research on SDN and energy efficiency in communication networks. In 2018 he received a fellowship from ``la Caixa'' Foundation to pursue his PhD at University of Vigo. His interests include SDN, Mobile Networks and Artificial Intelligence.
\end{IEEEbiography}

\begin{IEEEbiography}[{\includegraphics[width=1in,height=1.25in,clip,keepaspectratio]{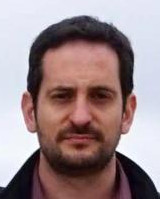}}]{Felipe Gil-Castiñeira} 
is currently an Associate Professor with the Department of Telematics Engineering, University of Vigo.
His  interests include wireless communication and core network technologies, multimedia communications, embedded systems, ubiquitous computing and the Internet of things. He has published over sixty papers in  international journals and conference proceedings. He has led several national and international R\&D projects. He holds two patents in mobile communications.
\end{IEEEbiography}

\begin{IEEEbiography}[{\includegraphics[width=1in,height=1.25in,clip,keepaspectratio]{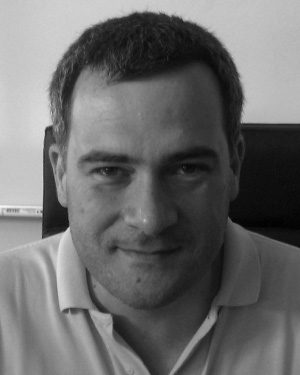}}]{Francisco Javier Gonzal\'ez-Casta\~no} is currently a Catedr\'atico de Universidad (Full Professor) with the Telematics Engineering Department, University of Vigo, Spain, where he leads the Information Technology Group. He has authored over 100 papers in international journals, in the fields of telecommunications and computer science, and has participated in several relevant national and international projects. He holds three U. S. patents.
\end{IEEEbiography}

\begin{IEEEbiography}[{\includegraphics[width=1in,height=1.25in,clip,keepaspectratio]{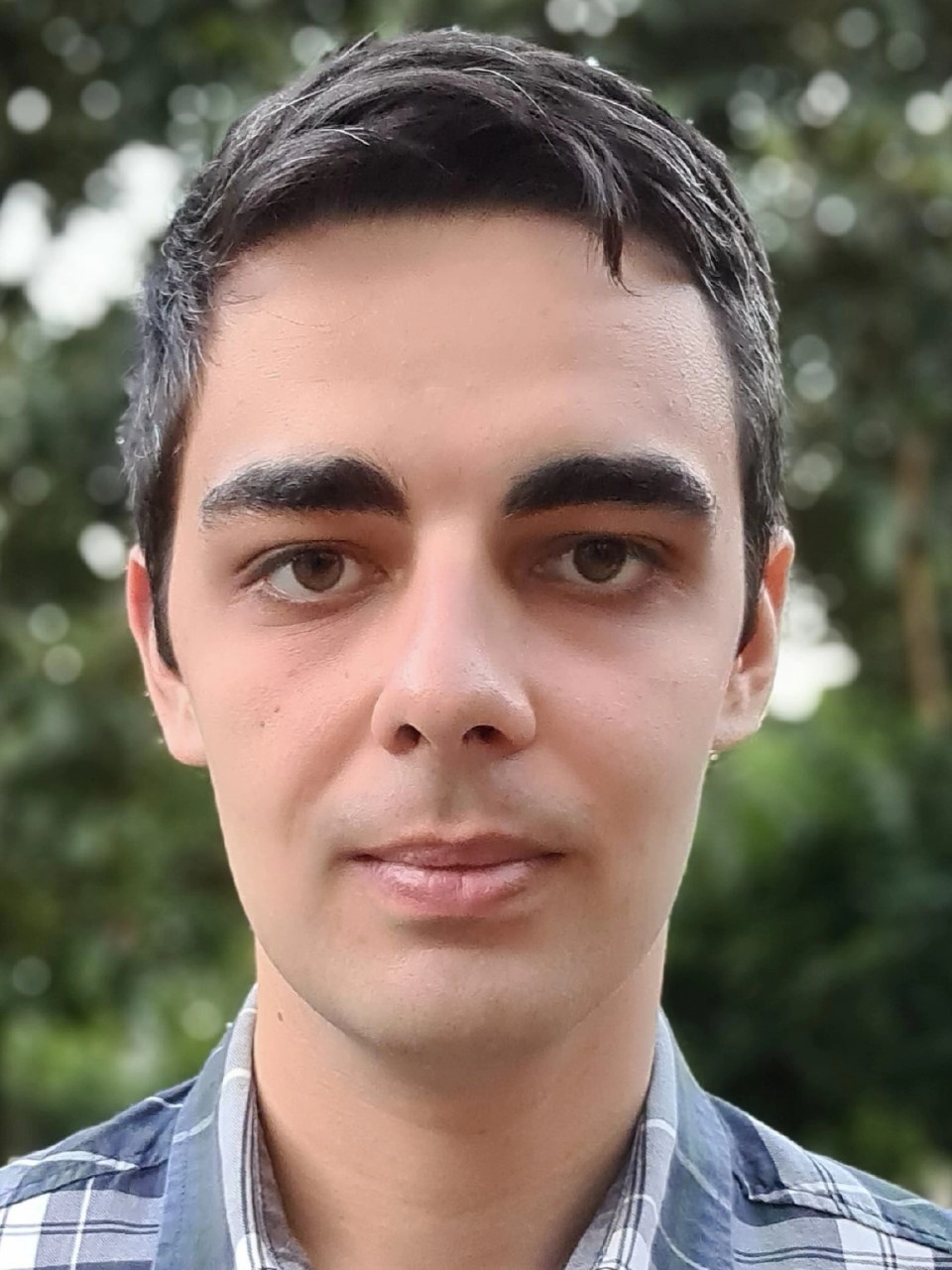}}]{David Candal-Ventureira} received the Bachelor's Degree in Telecommunication Engineering and the Master's Degree in Telecommunication Engineering from University of Vigo in 2016 and 2018, respectively. He is currently pursuing the Ph.D. Degree in Information and Communication Technologies. Since 2018 he has been working as a researcher in the Information Technologies Group, University of Vigo. His research interests include mobile and wireless networks and Artificial Intelligence.
\end{IEEEbiography}


\end{document}